\documentclass{aa}  

\usepackage{amsmath}
\usepackage{graphicx}
\usepackage{url}
\usepackage{txfonts}
\def\msun{{\rm ~M}_{\odot}}
\def\rsun{{\rm ~R}_{\odot}}
\def\zsun{{\rm ~Z}_{\odot}}
\def\gpy{{\rm ~Gpc}^{-3} {\rm ~yr}^{-1}}
\def\kms{{\rm ~km} {\rm ~s}^{-1}}

\usepackage{color}

\begin{document}

   \title{The origin of the first neutron star --- neutron star merger}

   \author{K. Belczynski\inst{1}\thanks{chrisbelczynski@gmail.com}
          \and
          A. Askar\inst{1}
          \and
          M. Arca-Sedda\inst{2}
          \and
          M. Chruslinska\inst{3}
          \and
          M. Donnari\inst{4}
          \and
          M. Giersz\inst{1}
          \and
          M. Benacquista\inst{5}
          \and
          R. Spurzem\inst{6}
          \and
          D. Jin\inst{5}
          \and 
          G. Wiktorowicz\inst{7,8}
          \and 
          D.Belloni\inst{1}
   }

   \institute{Nicolaus Copernicus Astronomical Center, Polish Academy of Sciences,
           ul. Bartycka 18, 00-716 Warsaw, Poland
         \and
           Zentrum fur Astronomie - Astronomisches Rechen-Institut, Weberstrabe 13, 
           D-69120 Heidelberg, Germany        
         \and   
           Department of Astrophysics/IMAPP, Radboud University, P.O. Box 9010, 6500 GL
           Nijmegen, The Netherlands
         \and
          Max-Planck-Institut f\"{u}r Astronomie, K\"{o}nigstuhl 17, 
          69117 Heidelberg, Germany
         \and
           Center for Gravitational Wave Astronomy, University of Texas Rio
           Grande Valley, Brownsville, TX, USA
         \and
           National Astronomical Observatories of China, Chinese Academy of Sciences
           NAOC/CAS, 20A Datun Rd., Chaoyang District, Beijing 100012, China
         \and
           National Astronomical Observatories, Chinese Academy of Sciences, 
           Beijing 100012, China
         \and
           School of Astronomy \& Space Science, University of the Chinese
           Academy of Sciences, Beijing 100012, China 
   }

   \date{Received Dec 3, 2017; accepted ???}

% \abstract{}{}{}{}{} 
% 5 {} token are mandatory
 
  \abstract
{ 
The first neutron star-neutron star (NS-NS) merger was discovered on August 
17, 2017 through gravitational waves (GW170817) and followed with electromagnetic 
observations~\citep{GW170817}. This merger was detected in an old elliptical 
galaxy with no recent star formation~\citep{Blanchard2017,Troja2017}. 
We perform a suite of numerical calculations to understand the formation mechanism 
of this merger. We probe three leading formation mechanisms of double compact 
objects: classical isolated binary star evolution, dynamical evolution in globular 
clusters and nuclear cluster formation to test whether they are likely to produce 
NS-NS mergers in old host galaxies. 
Our simulations with optimistic assumptions show current NS-NS merger rates 
at the level of $10^{-2}$ yr$^{-1}$ from binary stars, $5 \times 10^{-5}$ yr$^{-1}$ 
from globular clusters and $10^{-5}$ yr$^{-1}$ from nuclear clusters for all local 
elliptical galaxies (within $100$ Mpc$^3$). These models are thus in tension with 
the detection of GW170817 with an observed rate $1.5^{+3.2}_{-1.2}$ yr$^{-1}$ (per 
$100$ Mpc$^3$; LIGO/Virgo $90\%$ credible limits). 
Our results imply that either {\em (i)} the detection of GW170817 by LIGO/Virgo 
at their current sensitivity in an elliptical galaxy is a statistical coincidence; 
or that {\em (ii)} physics in at least one of our three models is incomplete in 
the context of  the evolution of stars that can form NS-NS mergers; or that 
{\em (iii)} another very efficient (unknown) formation channel with a long delay 
time between star formation and merger is at play.
}

   \keywords{Stars: massive -- Neutron-star physics -- Gravitational waves}

   \maketitle

\section{Introduction}

Double compact objects (NS-NS: neutron star --- neutron star systems; BH-NS:
black hole --- neutron star systems; BH-BH: black hole --- black hole systems) 
are considered to form along two major formation channels: isolated binary 
evolution in galactic fields~\citep{Tutukov1993,Lipunov1997,Voss2003,Belczynski2016b,
Stevenson2017} and dynamical evolution of stars in dense (e.g., globular) 
clusters~\citep{PortegiesZwart2004,Rodriguez2016a,Askar2017}. Each of these 
channels has one major variation that can be treated as a separate formation 
channel: homogeneous (rapid rotation) stellar evolution for isolated binaries
~\citep{Maeder1987,Yoon2005,deMink2009} and nuclear cluster evolution of stars 
with dynamical interactions~\citep{Antonini2016,Hoang2017,ArcaSedda2017}. 
The homogeneous evolution is claimed to work only for very massive stars 
($\gtrsim 30\msun$; e.g., BH progenitors) and not for stars that can produce 
neutron stars~\citep{Yoon2006,Mandel2016a,deMink2016,Marchant2016,Eldridge2016,
Woosley2016} and therefore we do not consider this channel in context of NS-NS 
formation.

For the remaining three channels (classical isolated binary evolution, globular 
cluster dynamics, and nuclear cluster dynamics) we perform estimates of the NS-NS merger 
rate. The estimates are done with a population synthesis method for the isolated binary
channel, with a Monte Carlo code that combines dynamical interactions with population 
synthesis for the globular cluster channel, and with a semi-analytical extrapolation of
globular cluster results to obtain an estimate for the nuclear cluster channel. 
Each estimate is self-consistent in its treatment of stellar evolution/dynamical 
evolution from star formation to NS-NS merger. However, the estimates are not fully 
consistent with each other as we choose different optimistic assumptions to increase 
the NS-NS merger rate within each channel. This allows us to independently assess the 
chance that GW170817 was formed along one of these channels, but it can not serve as a 
comparison between channels. 

Additionally, within each channel we use only a very limited knowledge of the effects 
of input physics on NS-NS merger rates. In the case of isolated binary evolution
our choice of input physics is based on a study of about 20 models with varying 
assumptions on the common envelope, Roche lobe overflow (RLOF) mass and angular 
momentum loss, and natal kicks~\citep{Chruslinska2018}. In the case of the dynamical
channel our choice of input physics is based on previous simulations of 
$\sim 2000$ globular cluster models with varying initial (mass, stellar density, and 
binarity) cluster parameters~\citep{Askar2017}. In the case of the nuclear cluster channel 
we test two major models of nuclear cluster formation in the context of NS-NS merger 
production. For each channel we present only one model with specific input
physics that tends to increase the NS-NS merger rate. The 
various models that provide the basis for our choice of input physics are only a 
small representation of existing possibilities within the multi-dimensional 
parameter space of uncertain evolutionary and dynamical parameters. Our main
goal is to show what are the highest NS-NS merger rates that are attainable 
with {\em currently} tested models and contrast them with the LIGO/Virgo 
detection. Future studies will hopefully gain extra momentum to test broad 
regions of parameter space if we can demonstrate here that LIGO/Virgo estimate 
of merger rate cannot be reproduced with current set of models. This is all 
based on the assumption that LIGO/Virgo single detection {\em is not} a 
statistical coincidence.

For all three channels we need as an input star forming mass that may have
potentially contributed to the formation of GW170817 in an old host galaxy.  
Advanced LIGO/Virgo was sensitive to NS-NS mergers to $\lesssim 100$ Mpc. 
The host of GW170817, NGC 4993, is an early type galaxy with sub-solar metallicity
and with virtually no current star formation. The peak of star formation was estimated
at $\gtrsim 10$ Gyr (with exponential decay afterwards; \cite{Blanchard2017}), 
or last estimated to take place $t_{\rm sf}=3$--$7$ Gyr ago~\citep{Troja2017}.  
For our ``realistic'' estimates we assume all early type galaxies within $100$ Mpc$^3$ 
around Earth formed all stars $t_{\rm sf}=5$ Gyr ago. We also allow for two variations, 
one with $t_{\rm sf}=10$ Gyr (``pessimistic'') and another with $t_{\rm sf}=1$ Gyr 
(``optimistic''). Using the Illustris cosmological simulation we estimate number of 
local ($z=0$) elliptical galaxies to be $N_{\rm ell}=65,821$ within $100$ Mpc$^3$. 
Stellar mass contained in these elliptical galaxies is 
$M_{\rm ell,tot}=1.1 \times 10^{14} \msun$ (see App.~\ref{sec.app0}). 
The Illustris cosmological simulation shows a remarkably good agreement with the 
overall observed properties of galaxies at low redshift~\citep{Vogelsberger2014}. 
It also provides a good representation of the galaxy stellar mass function and the 
evolution of star formation across cosmic times~\citep{Genel2014} and a reasonable 
variety of galaxy morphology and colors~\citep{Snyder2015,Torrey2015}.
For all estimates we assume sub-solar metallicity $Z=0.01$. 

In practical terms, in our evolutionary simulations (see Sec.~\ref{sec.bin}, 
~\ref{sec.gc}, ~\ref{sec.nc}) we assume that the entire considered mass ($M_{\rm
ell,tot}$) forms stars at one specific time (delta function SFR) in the past. 
And then we test whether this amount of stars can form enough NS-NS mergers 
to account for detection of GW170817 with the current LIGO/Virgo sensitivity. 
The LIGO/Virgo estimated rate of NS-NS mergers, based on this single detection 
is at the level of $1.5$ yr$^{-1}$ within $100$ Mpc$^3$ (with $90\%$ credible 
range: $0.3$--$4.7$ yr$^{-1}$ \cite{GW170817}).

\section{Classical Isolated Binary Evolution}
\label{sec.bin}

\begin{table}
\caption{Local NS-NS merger rates [yr$^{-1}$] (within $100$ Mpc$^3$).} 
\centering          
\begin{tabular}{c| c c c}     
\hline\hline       
Model                        & pessimistic &   realistic  & optimistic \\ 
\hline 
\hline
LIGO/Virgo\tablefootmark{a}  &      $0.3$         &      $1.5$           & $4.7$  \\ 
\hline
                             &                    &                      & \\
classical binaries           & $8 \times 10^{-3}$ &  $1 \times 10^{-2}$  & $5 \times 10^{-2}$ \\  
globular clusters            & $2 \times 10^{-5}$ &  $5 \times 10^{-5}$  & $5 \times 10^{-4}$ \\                   
nuclear clusters             & $7 \times 10^{-6}$ &  $1 \times 10^{-5}$  & $1 \times 10^{-4}$ \\
\hline      
\hline
\end{tabular}
\label{tab.rates}
\tablefoot{ \\
\tablefootmark{a}{
The LIGO/Virgo estimate ($1540^{+3200}_{-1220} \gpy$) rescaled by $0.001$ to
show merger rate within $100$ Mpc$^3$.}\\
}
\end{table}

\subsection{Overall Description}

We use the population synthesis code {\tt StarTrack}~\citep{Belczynski2002,
Belczynski2008a} to evolve stars in isolation (in galactic fields) without taking 
into account effects of rapid rotation to generate a population of NS-NS 
binaries. We start with $M_{\rm ell,tot}=1.1 \times 10^{14} \msun$ stars with 
initial properties guided by recent observations~\citep{Sana2012} and assume 
$100\%$ binary fraction. Some of the binary configurations leading to mass 
transfers and common envelope, survive through supernovae that may 
or may not disrupt them to form close NS-NS binaries that merge 
within a Hubble time. 
We choose an evolutionary model in which we adopt our standard input physics 
(see Sec.~\ref{sec.app1}) with the additional assumptions that each stable RLOF 
is fully conservative (i.e., no mass is lost from a binary) and that stars on 
Hertzsprung gap are allowed to initiate and survive common envelope (CE). This 
tends to increase NS-NS merger rates in classical isolated binary 
evolution~\citep{Chruslinska2018}. 
The typical formation of a NS-NS system involves a common sequence: RLOF (from primary), CE 
(from secondary), and RLOF (from secondary). However with the above assumptions 
progenitors evolve typically through a different sequence: RLOF (from primary), 
CE (from secondary), and CE (from secondary). Two CE events lead to formation
of very close NS-NS systems increasing merger rates. During conservative
RLOF (from the primary) the secondary star becomes more massive than in the non-conservative 
case. This makes the secondary envelope, when it expands after the main sequence, more 
massive and the ensuing CE leads to a more drastic orbital decay. After the first 
CE, the secondary exposed helium core expands and initiates a second CE (rather than
RLOF as it is more massive than in the non-conservative case) right after the core-He 
burning phase (helium Hertzsprung gap). In general, keeping more mass 
(conservative RLOF) in binary systems allows NS-NS mergers to occur from lower 
mass stars (increasing rates thanks to the IMF) and allowing for a more liberal 
application and survival of CE also increases rates~\citep{Dominik2012}.

\begin{figure}   
\hspace*{-0.3cm}   
\includegraphics[width=9.2cm]{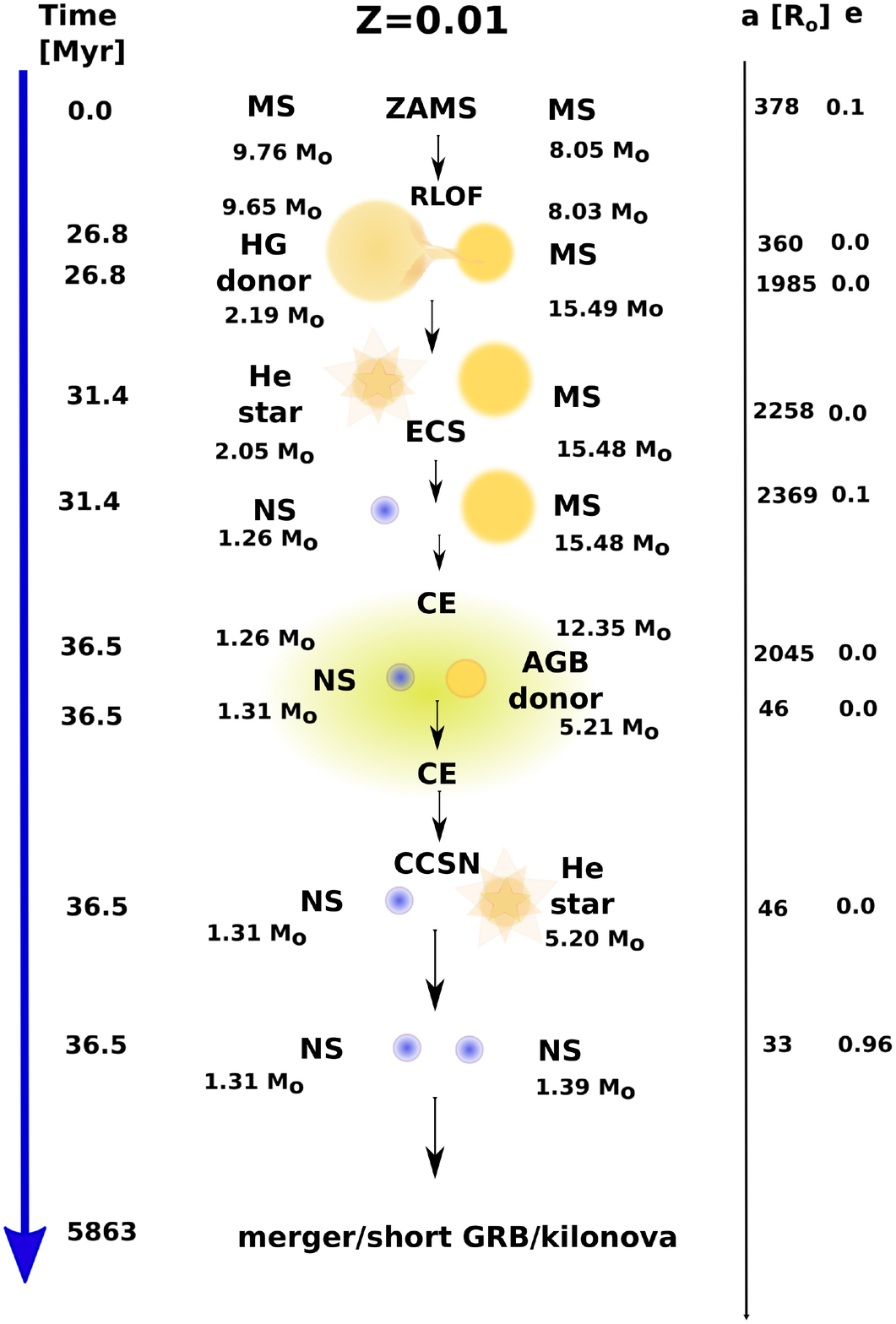}  
\caption{    
Example of the formation of a NS-NS merger similar to GW170817 in the 
classical isolated binary evolution channel.  
}
\label{fig.nsns1}
\end{figure}

Our simulations show that it is possible to form NS-NS mergers in old
elliptical galaxies. An example is shown in Figure~\ref{fig.nsns1} (a detailed 
description of this evolutionary example is given in Sec.~\ref{sec.app1b}).  
However, the predicted current merger rate of NS-NS systems from all ellipticals 
within $100$ Mpc$^3$ is low: $R_{\rm nsns}=0.01$ yr$^{-1}$ for $t_{\rm sf}=5$ 
Gyr. For comparison, the LIGO/Virgo estimated rate of NS-NS mergers is
$1.5$ yr$^{-1}$ within $100$ Mpc$^3$. Our predicted rates decrease for the older 
star formation (see Tab.~\ref{tab.rates}). Rates can be as high as 
$R_{\rm nsns} \sim 0.05$ yr$^{-1}$ for $t_{\rm sf}=1$ Gyr, but this is still 
well below the LIGO/Virgo low estimate. Additionally, it is very unlikely 
that majority of elliptical galaxies have stellar populations as young as 
$t_{\rm sf} \lesssim 1$ Gyr. 

\begin{figure}
\hspace*{-0.3cm}
\includegraphics[width=9.2cm]{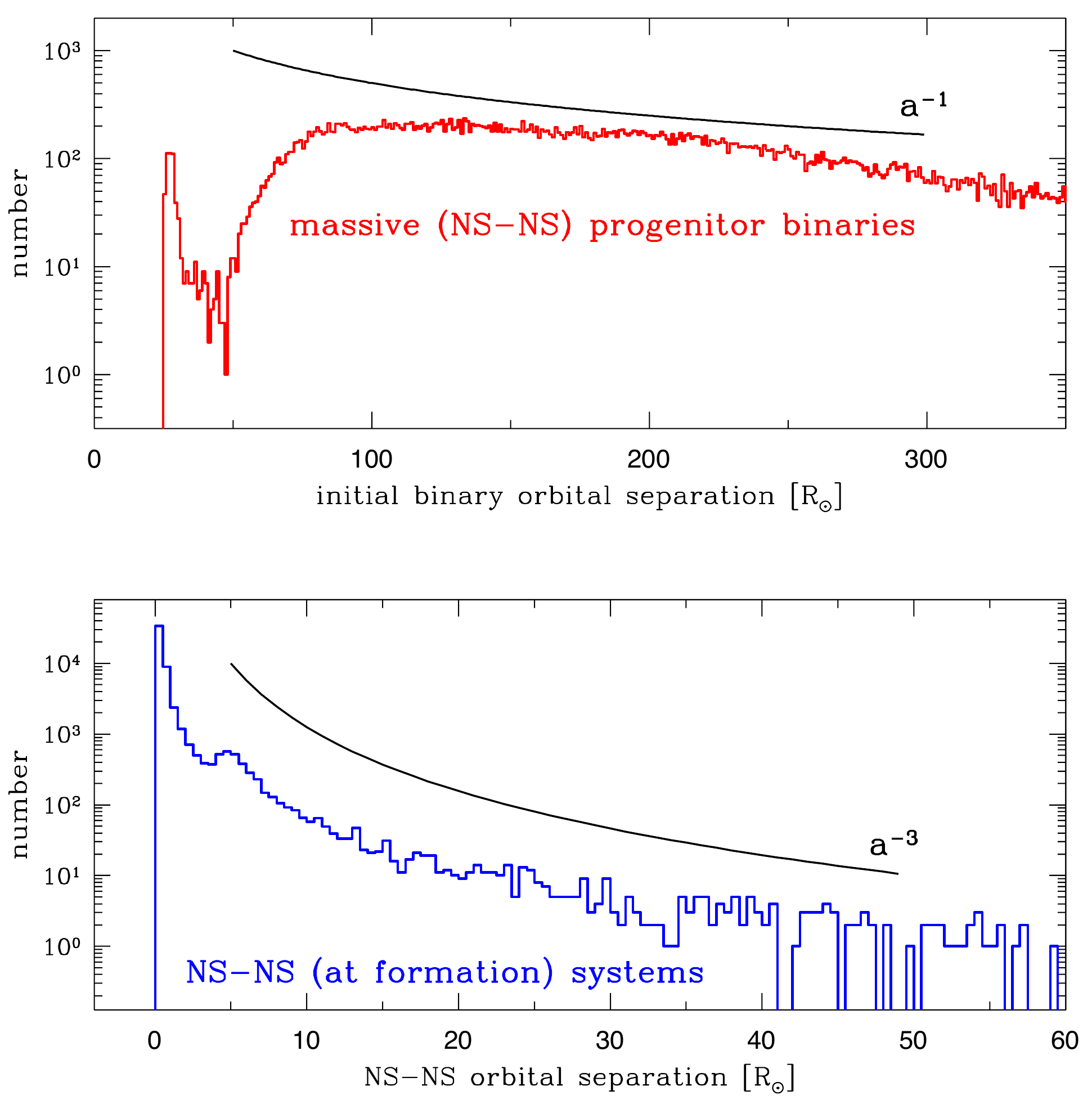}
\caption{
Initial orbital separation of binaries that are progenitors of NS-NS
mergers; note that the distribution is close to $a^{-1}$ (top). After binary
evolution (mass transfers, supernovae, CE) close NS-NS systems form with
much smaller orbital separations, and their orbital separation distribution
may be approximated by a steep power-law: $a^{-3}$ (bottom).
}
\label{fig.orbit}
\end{figure}

The decrease of the merger rate with time is due to typically short delay 
times from star formation to NS-NS merger. Initial orbital separations are 
observed to be steep power-laws for massive binaries (e.g., $\propto a^{-1}$ 
-- $a^{-2}$; \cite{Kobulnicky2014,Sana2012}). Complex evolutionary processes 
(mass exchanges, supernovae natal kicks and mass loss, CE evolution) are bound 
to modify orbital separations before NS-NS formation. However, NSs form from 
stars in a relatively narrow mass range and typically NS-NS merger formation 
is dominated by one specific evolutionary sequence~\citep{Dominik2012}. The 
net effect of the evolutionary processes is rather similar for most of the  
NS-NS progenitors, decreasing initial separations to smaller values (CE 
evolution) and (approximately) steepening the shape of the orbital separation 
distribution (see Fig.~\ref{fig.orbit}). For NS-NS orbital separations that 
are a steep power-law ($\propto a^{-3}$) the convolution with the gravitational 
radiation emission orbital decay timescale ($\propto a^4$; \cite{Peters1964}) 
results in a power-law delay time distribution $\propto t^{-1.5}$. The delay 
time scales as $a^{-3} (da/dt)_{\rm GR} \propto t^{-3/4} d(t^{1/4})/dt \propto t^{-1.5}$. 
For our particular model ($Z=0.01$) the majority of NS-NS mergers occur within 
$1$ Gyr of star formation: $97\%$ (see Fig.~\ref{fig.tdel}). Although the delay 
time distributions differ for other evolutionary models and other metallicities, 
they are still steep power-laws~\citep{Dominik2012}. This implies that NS-NS 
mergers are typically predicted in young stellar populations (e.g., in 
starbursts or spirals with the ongoing/recent star formation), although some 
fraction is still to be expected even in galaxies with no star formation (e.g., 
ellipticals).

\subsection{Details of Calculations}
\label{sec.app1}

Our evolutionary modeling is performed with the {\tt StarTrack} Monte Carlo 
population synthesis code~\citep{Belczynski2002}. In particular, we incorporate 
a calibrated treatment of tidal interactions in close binaries~\citep{Belczynski2008a},
a physical measure of the common envelope (CE) binding energy~\citep{Dominik2012,Xu2010},
and a rapid explosion supernova model that reproduces the observed mass gap between
neutron stars and black holes~\citep{Belczynski2012}. Our updated compact object 
mass spectrum covers a wide range of NS masses ($M_{\rm NS}=1.1$--$2.5\msun$; 
\cite{Fryer2012}). Neutron stars are formed from single stars with initial mass 
range $M_{\rm zams}=7.4$--$7.9\msun$ in electron capture supernovae and in range 
$M_{\rm zams}=7.9$--$21.0\msun$ in core-collapse supernovae for the sub-solar 
metallicity considered in our study $Z=0.01$. These ranges are subject to change 
due to effects of mass accretion and loss in binary evolution. In particular, 
even stars as massive as $M_{\rm zams}\sim 100\msun$ may form NSs in binaries 
while losing most of their mass in case A RLOF~\citep{Belczynski2008c}. 

Based on our previous modeling~\citep{Chruslinska2018} we consider one specific 
variation of the binary evolution input physics that tends to increase NS-NS merger 
rates. We allow for Hertzsprung gap (HG) stars to initiate and survive common 
envelope (CE) evolution. This is an optimistic assumption, since these stars may 
not initiate the CE evolution, or may not survive as a binary if CE does happen
~\citep{Belczynski2007,Pavlovskii2015}. Note that CE is a major evolutionary
process needed for the formation of double compact object mergers in our
evolutionary framework~\citep{Belczynski2002}. During CE we adopt a standard
energy-based formalism to calculate the orbital decay~\citep{Webbink1984} and we
assume that $100\%$ of orbital energy is used to eject the envelope and envelope 
binding energy is obtained from detailed calculation of stellar structure with 
partial inclusion of ionization energy~\citep{Dominik2012}. 
During stable Roche lobe overflow (RLOF) we assume that mass transfer is fully 
conservative and no angular momentum is lost from the binary. This particular 
assumption allows for rather effective NS-NS binary formation. For NS formation 
in electron-capture supernova (ECS; \cite{Miyaji1980,Podsiadlowski2004}) we 
assume that there is no associated natal kick. However, some small natal kick 
velocity ($\lesssim 50 \kms$) may result from such explosions 
~\citep{Dessart2006,Jones2013,Schwab2015}. We assume that NS forming in iron 
core-collapse SNe receive natal kicks with velocity components drawn from a 
$1$-D Maxwellian distribution with rms $\sigma_{0}=265\kms$~\citep{Hobbs2005}.
The magnitude of the kick is further decreased by the amount of fallback 
estimated for each NS at its formation (important only for the heaviest NSs; 
\cite{Fryer2012}). Natal kicks are assumed to have random direction. 
Lowering the iron core-collapse supernovae natal kicks leads to only moderate 
increase in the predicted NS-NS merger rates (e.g., using $\sigma=\sigma_{0}/2$ 
would increase the rates by a factor of $\lesssim 1.5$; \cite{Chruslinska2017}). 
For massive O/B stars that are NS progenitors we apply mass and metallicity
dependent wind mass loss~\citep{Vink2001}, while for naked helium stars we
apply combination of wind rate estimates that take into account Wolf-Rayet 
stellar wind clumping~\citep{Hamann1998}, and wind metallicity-dependence 
for Wolf-Rayet stars ($\propto (Z/\zsun)^{0.86}$; \cite{Vink2005}).

Our model (for all NS progenitors) is computed with initial distributions 
of orbital periods ($\propto (\log P)^{-0.55}$), eccentricities ($\propto e^{-0.42}$), 
and mass ratios ($\propto q^{0}$) appropriate for massive O/B stars~\citep{Sana2012}.
We adopt an initial mass function that is close to flat for low mass stars
($\propto M^{-1.3}$ for $0.08 \leq M<0.5\msun$ and $\propto M^{-2.2}$ for 
$0.5 \leq M<1.0\msun$) and top heavy for massive stars ($\propto M^{-2.3}$ 
for $1.0 \leq M \leq 150\msun$), as guided by recent observations
~\citep{Bastian2010}.

\subsection{Example of Calculations}
\label{sec.app1b}

Example of NS-NS merger in old host galaxy formed in our model of classical 
isolated binary evolution is shown in Figure~\ref{fig.nsns1}. The evolution
begins with two relatively low-mass stars ($M_{\rm zams,a}=9.76\msun$ and
$M_{\rm zams,b}=8.05\msun$) with moderately sub-solar ($Z=0.01$) metallicity,
placed on a wide ($a=378\rsun$) and almost circular orbit ($e=0.1$).

Primary (initially more massive) star evolves off the Main Sequence (MS) and 
during the subsequent Hertzsprung gap evolution initiates a stable RLOF, 
transferring its entire H-rich envelope to the secondary star. 
In this process the primary turns into a low mass naked helium star while 
the secondary becomes a massive (rejuvenated) MS star ($M_{\rm a}=2.19\msun$ 
and $M_{\rm b}=15.49\msun$; note the mass ratio reversal). The orbit circularizes 
($e=0$) and expands in response to this fully conservative mass transfer 
($a=1985\rsun$). During the late stages of its evolution the primary expands to 
become a helium-rich giant ($R \sim 100 \rsun$) and loses part of its envelope 
in stellar winds, reducing its mass to $M_{\rm a}=2.05\msun$ .
This leads to a moderate orbital expansion ($a=2258\rsun$). Finally, the primary 
forms a low-mass oxygen-neon-magnesium core that collapses due to electron 
capture processes and leads to electron-capture supernova. 
We assume that a relatively lightweight ($M_{\rm a}=1.26\msun$) neutron star 
with no natal kick is formed in this process. However, a supernova mass 
ejection and neutrino emission (both assumed to be fully symmetric in ECS case) 
still affect the orbital parameters of the binary, increasing the orbital 
separation ($a=2369$) and eccentricity ($e=0.1$). The first NS forms $t=31.4$ 
Myr after the beginning of its evolution on Zero Age Main Sequence (ZAMS). 

As the massive secondary star evolves off MS, it expands and is subject to
significant wind mass loss. During asymptotic giant branch (AGB) evolution 
its mass decreases ($M_{\rm b}=12.35\msun$), while its size increases enough
($R \sim 800 \rsun$) to start the second RLOF. Tidal forces (spinning up the 
expanding secondary) circularize the orbit ($e=0$) and reduce the orbital 
separation ($a=2045$). Due to high mass ratio (of AGB secondary to primary NS) 
at this time and response of convective envelope of the secondary to mass loss 
this RLOF is dynamically unstable and leads to CE evolution. 
CE leads to severe reduction of the orbital size ($a=46\rsun$) of the system.
The secondary is stripped of its entire H-rich envelope and becomes a massive 
helium star ($M_{\rm b}=5.21\msun$). We allow the primary NS to accrete 
during CE at $10\%$ Bondi-Hoyle rate~\citep{MacLeod2017} and as a result the 
NS increases its mass ($M_{\rm a}=1.31\msun$). At time $t=36.5$ Myr secondary 
star forms an iron core that collapses and the star explodes as Type Ib 
supernova.
This supernova results in a significant mass of ejecta and we assume that 
$10\%$ gravitational mass is lost in neutrino emission. We calculate the 
secondary NS mass ($M_{\rm a}=1.39\msun$). The explosion also leads to a 
moderately high natal kick (3D magnitude: $83\kms$; either due to asymmetric 
mass ejection~\citep{Janka1994}; asymmetric neutrino emission~\citep{Fryer2006b}; 
or the combination of both) gained by the newly formed NS and the orbit becomes 
highly eccentric ($a=33\rsun$ and $e=0.96$).

The orbital parameters of the resulting NS-NS binary lead to long delay time of 
$5.8$ Gyr (Peters 1964). 
Such a system might have formed long ago in NGC 4993 and it would have merged
close to the present time, allowing for the detection of gravitational waves
similar to GW170817 and would be accompanied by short GRB and kilonova. However, 
we note that this is not a typical NS-NS binary found in our simulations. The 
majority of merging NS-NS systems forms with short delay times ($t\lesssim 1$ Gyr; 
$\propto t^{-1}$ or somewhat steeper) and follows other formation channels
~\citep{Chruslinska2018}.

\section{Globular Cluster Dynamics}
\label{sec.gc}

\subsection{Overall Description}

We use the {\tt MOCCA} code~\citep{Giersz2013} to compute a suite of globular 
cluster (GC) models with updated prescriptions for binary and stellar evolution. 
For all NS progenitors we have adopted a standard IMF~\citep{Kroupa2001} and 
evolved stars with initial properties (orbital periods: $\propto (\log P)^{-0.55}$, 
eccentricities: $\propto e^{-0.42}$, mass ratios: $\propto q^{0}$ 
as observed for massive O/B stars~\citep{Sana2012}.
The difference between this IMF and the one used for the field calculations
in Sec. \ref{sec.app1} is that stars between $0.5$ and $1.0 \msun$ have a
power-law index of $-2.3$ instead of $-2.2$ and maximum initial ZAMS mass is
$100 \msun$ instead of $150 \msun$.
The evolution of 27 GC models was simulated to $15$ Gyr and the models span a 
range of initial parameters including cluster mass, size, and binary 
fraction (see Sec.~\ref{sec.app2}). All GC models were initially isotropic
~\citet{King1966} models with central concentration parameter ($\rm W_{0}$) value
of 6. The current Milky Way mass fraction in GCs is $\lesssim 0.002$, and GCs 
were initially $\sim 5$ times more massive~\citep{Webb2015}. We assume that
the same holds for elliptical galaxies and that the fraction of stellar mass of 
elliptical galaxies ($M_{\rm ell,tot}$) found in GCs is $0.01$. This gives 
total initial mass $M_{\rm gc,tot}=1.1 \times 10^{12} \msun$ in all GCs found 
in all ellipticals within $100$ Mpc$^3$. 
Based on observations of $48,000$ globular clusters in $7$ supergiant cluster 
galaxies~\citep{Harris2014} we adopt a log-normal initial GC mass distribution
with mean $M_{\rm gc,ave}=1.8 \times 10^6\msun$ and $\sigma=0.5$ in the range of 
plausible initial GC masses: $M_{\rm gc}=5 \times 10^{3}$--$5 \times 10^{7}\msun$. 
This corresponds to $N_{\rm gc}=6.4 \times 10^5$ GCs in ellipticals within $100$ 
Mpc$^3$.  

\begin{figure}   
\hspace*{-0.3cm}   
\includegraphics[width=9.2cm]{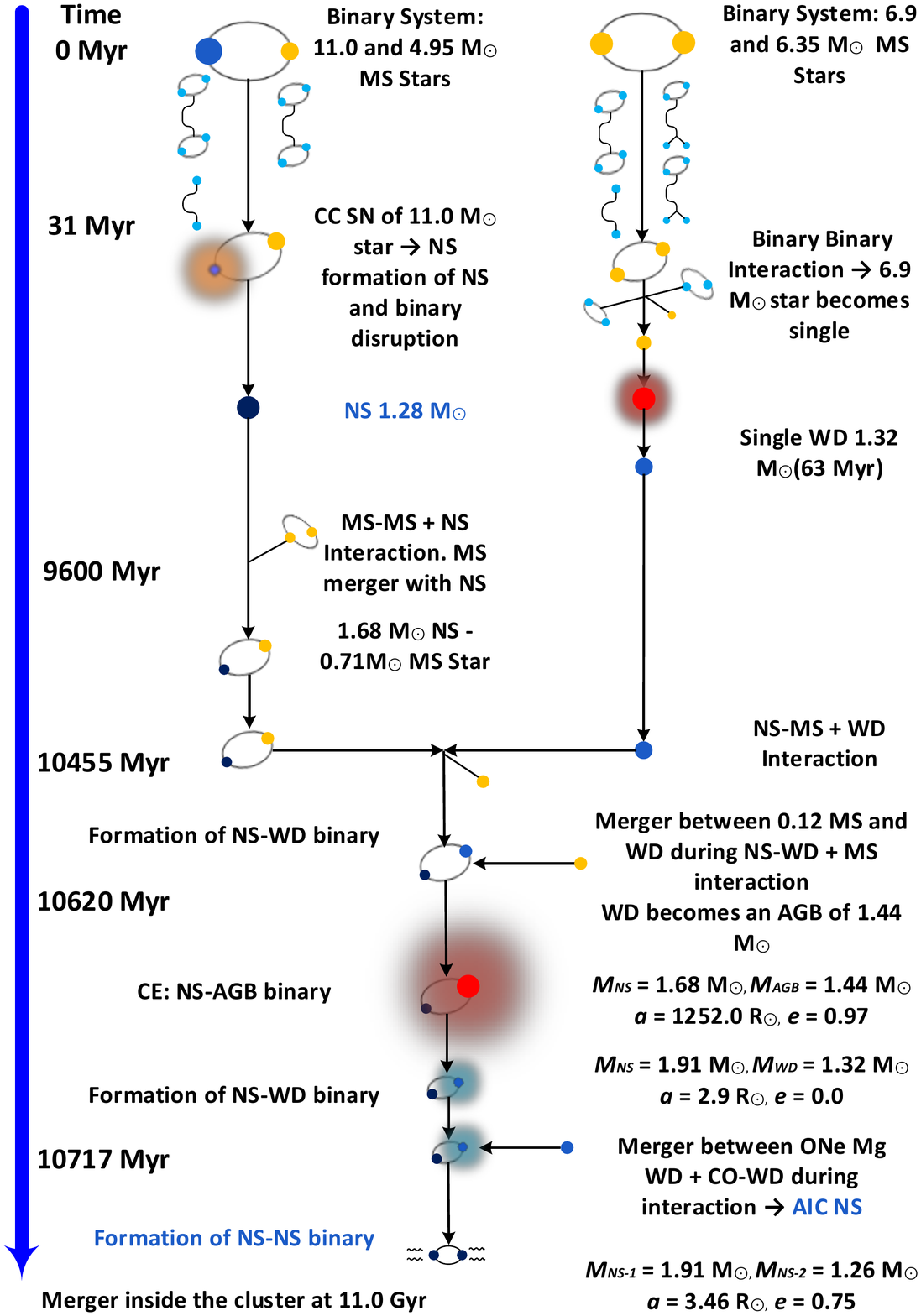}  
\caption{    
Example of the formation of a NS-NS merger similar to GW170817 in the 
globular cluster dynamical channel.  
}
\label{fig.nsns2}
\end{figure}

We find that our GC models can dynamically produce NS-NS binaries that 
will coalesce within a Hubble time (e.g., Fig.~\ref{fig.nsns2}). The number of
coalescing NS-NS binaries in a GC cluster model depends strongly on the 
initial parameters and in particular on the initial cluster mass: 
$N_{\rm nsns} = 0.001 \times (M_{\rm gc}/\msun)^{0.55}$. This relation shows 
that a GC with an initial mass of $7 \times 10^{5} \msun$ can at best produce 
$2$ NS-NS mergers. The predicted GC NS-NS merger rate from all ellipticals within 
$100$ Mpc$^3$ for $t_{\rm sf}= 5$ Gyr is $R_{\rm nsns}=5 \times 10^{-5}$ yr$^{-1}$. 
Our estimated rate increases, the younger the GC is. In particular, 
$R_{\rm nsns}=2 \times 10^{-5}$ yr$^{-1}$ for $t_{\rm sf}= 10$ Gyr, and 
$R_{\rm nsns}= 5 \times 10^{-4}$ yr$^{-1}$ for $t_{\rm sf}= 1$ Gyr.

Although there are differences in the initial setup and a few prescriptions 
for physical processes involved in isolated binary evolution are different, 
the most optimistic GC merger rates of NS-NS binaries are $\sim 2$ orders of 
magnitude lower than the rates from classical isolated binary evolution (see 
Tab.~\ref{tab.rates}).  
The main reason for this is the fact that the stellar mass in GCs is much lower 
($0.01 M_{\rm ell,tot}$) than in the field ($M_{\rm ell,tot}$). Although a GC 
model can dynamically form many BH-BH binaries during its dynamical evolution, the 
number of dynamically formed NS-NS binaries is much lower. Initially dense GC models 
with large escape velocities that would retain a high fraction of NS ($\gtrsim 0.5$) 
are also more likely to retain a high fraction of BHs.
These retained BHs can segregate to the center of the globular cluster,
forming a subsystem comprising of single and binary black holes that will 
provide energy to the surrounding stars and support the evolution of the GC
~\citep{Breen2013a,Breen2013b}. Many recent GC simulations
\citep{Morscher2013,Sippel2013,Heggie2014,Morscher2015,Wang2016,Peuten2016,
Rodriguez2016-dr,ArcaSedda2018a,ArcaSedda2018b} have shown that depending on the
initial GC model, such a subsystem of BHs can survive up to a Hubble time and
dominate the central dynamics of the GC. The presence of a sizeable BH subsystem
prevents segregation of NSs to the GC center which inhibits the formation of NS-NS
binaries through strong dynamical interactions.
 For a moderately dense GC model (with initial 
half mass radius of $r_{\rm h}=1.2$ pc and galactic tidal radius of $r_{\rm t}=60$ 
pc), we note a peak of NS-NS mergers originating from primordial binaries within the 
first Gyr of GC evolution, then there is a long period ($1$--$10$ Gyr) of low NS-NS 
merger rate (primordial NS-NS mergers dying off, while dynamical NS-NS mergers are 
just beginning to appear), and finally dynamical mergers are beginning to peak at 
late times ($>10$ Gyr after the star formation). This late time corresponds to the 
depletion of the BH subsystem and the subsequent core collapse of the GC. NS-NS 
mergers are found to take place either within GCs ($\sim 35\%$) or after ejection 
from their host GCs ($\sim 65\%$).

\subsection{Details of Calculations}
\label{sec.app2}

Results are obtained using the 
{\tt MOCCA} (MOnte Carlo Cluster simulAtor) code for star cluster simulations 
\citep[see][and reference therein]{Giersz2013,Hypki2013}. The code treats 
dynamical relaxation of stars and binary systems in spherically symmetric star 
clusters using the Monte Carlo method for stellar dynamics developed by 
\citet{Henon1971} which was further improved by \citet{Stodolkiewicz1986} 
and \citet{Giersz1998}. For strong dynamical interactions between binary 
systems and binaries and single stars, {\tt MOCCA} uses the {\tt FEWBODY} \citep{Fregeau2004} 
for simulating small-N gravitational dynamics. For basic stellar and binary 
evolution routines, the {\tt MOCCA} code uses prescriptions from the {\tt SSE/BSE} 
code \citep{Hurley2000,Hurley2002} with updates which include formation and proper 
treatment of NSs via ECS and accretion induced collapse (AIC). Other changes 
were also made to SSE/BSE prescriptions (Belloni et al. submitted) based on 
recent updates to stellar/binary evolution routines in the latest version of 
{\tt NBODY6} \footnote{A short summary of few of these updates is available at
\url{ftp://ftp.ast.cam.ac.uk/pub/sverre/nbody6/README_SSE}} and {\tt StarTrack}.

The evolution of 27 GC models was simulated up to 15 Gyr and these models 
spanned a range of initial parameters which include cluster mass (from 
$6.25 \times 10^{4} \msun $ up to $1.3 \times 10^{6} \msun $, size 
(half-mass radii of $1.2$ and $2.4$ pc), Galactocentric radius (2.5 kpc to 10.8 kpc), 
binary fraction ($10\%$ and $95\%$) and metallicity ($Z=0.01$ and $Z=0.002$).
All models were initially non-segregated isotropic \citet{King1966} models 
with central concentration parameter ($\rm W_{0}$) value of $6$ and each model 
also has a two component IMF given by \citet{Kroupa2001} with stellar 
masses in the range $0.08$ and $100.0 \msun $.
For models with $95\%$ initial binaries, the semi-major axis, eccentricity and
mass ratio distributions are given by \citet{Belloni2017a,Kroupa1995}, and
in particular for binaries with O/B stars \citet{Sana2012} distributions are
used. For models with $10\%$ binaries, we used a uniform mass ratio distribution, 
a uniform distribution in the logarithm of the semi-major axis and a thermal 
eccentricity distribution. BH natal kicks were 
computed using the mass fallback prescription of \citet{Belczynski2002}.
The prescription for BH natal kicks is different than the one used for the
field calculations. We do not expect that a different fallback prescription for
BHs will drastically change the results for NS-NS binaries that originate from
GCs. Having high natal kicks for BHs could be helpful in preventing the formation 
of a BH subsystem which may lead to more centrally segregated NSs, some of which
may dynamically form NS-NS binaries. However, giving large natal kicks to BHs
would undermine results estimating BH-BH binaries produced in GCs
~\citep{Rodriguez2015,Rodriguez2016a,Rodriguez2016b,Rodriguez2016c,Askar2017,Park2017}. 
NSs formed in iron core-collapse  were given natal kicks with a Maxwellian 
distribution with $\sigma=100\kms$ or zero natal kicks to have a higher
NS retention factor and for checking the maximum contribution GC NS-NS binaries 
could have to merger rates. These values are significantly lower than NS natal 
kick values that are typically used in GC simulations. Based on proper motion 
estimates of pulsars in our Galaxy \citep{Hobbs2005}, NS natal kicks are usually 
given by a Maxwellian distribution with $\sigma=265\kms$. In all models, NSs 
forming via ECS or AIC were given zero natal kicks at birth.  

For a small sub-sample of models, additional runs were also simulated in which 
common envelope evolution parameters were changed in order check the influence 
of CE on NS-NS formation. We either calculated the binding energy parameter 
$\lambda$ for the giant in CE (by setting $\lambda$ to 0.0 in BSE) or fixed 
the value to be $\lambda=0.05$. For all runs, the $\alpha$ parameter, which is the 
fraction of orbital energy used to unbound the envelope was set 
to $1$ \citep[see Section 3.2 in][for details about CE in {\tt BSE}]{Belloni2017b}. It 
is important to stress that the purpose of this study was to obtain an 
order of magnitude estimate for the rate and for this reason, a limited number 
of GC models with assumptions conducive to formation of NS-NS binaries (low and 
zero natal kicks) were simulated. Future works, will cover a more detailed 
parameter space in combination with better constraints for the galactic 
environment. 

\begin{figure}   
\hspace*{-0.3cm}   
\includegraphics[width=9.2cm]{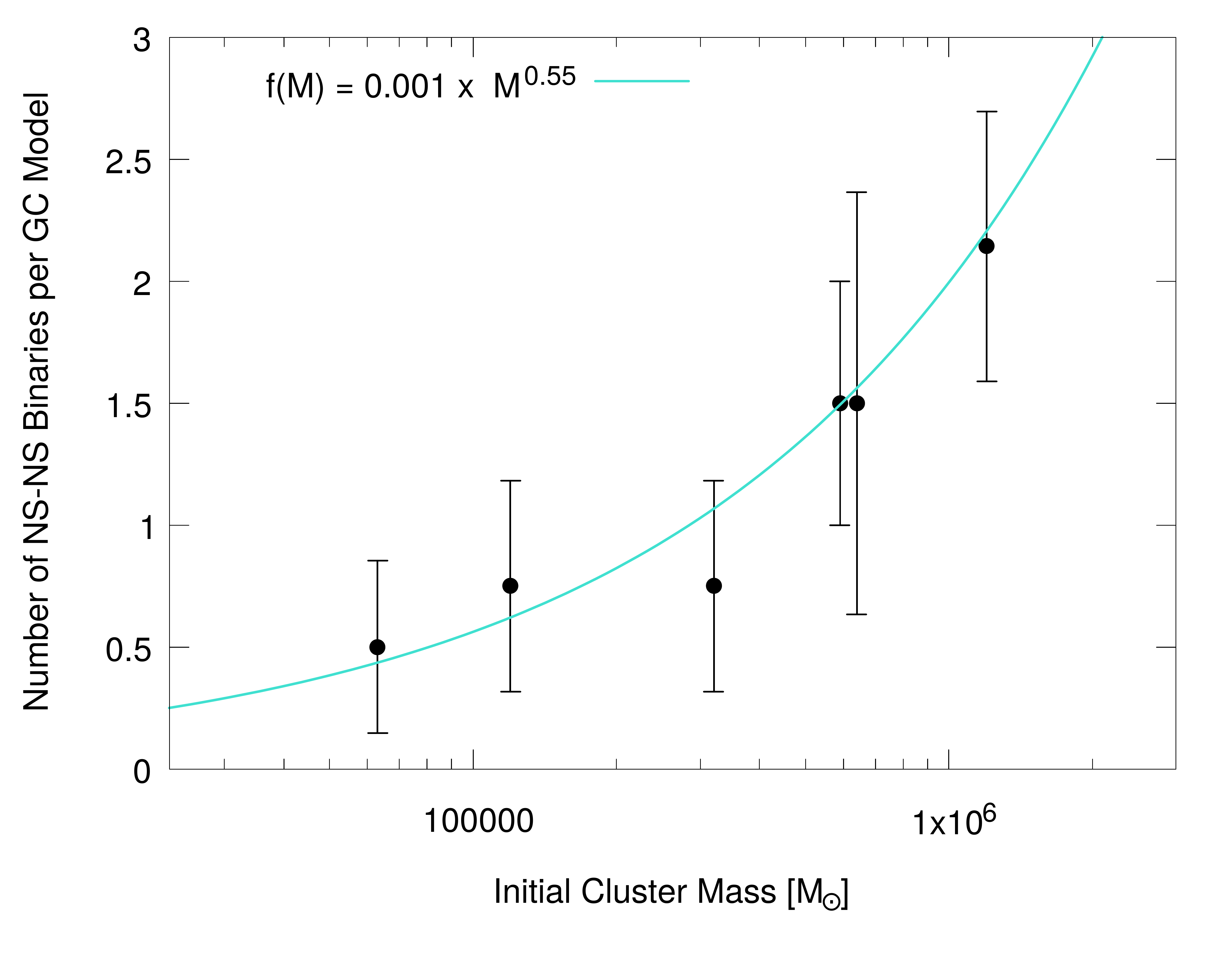}  
\caption{    
The figure shows the correlation between initial mass of GCs and the estimated 
number of NS-NS binaries such a cluster could produce. The error bars represent
the Poisson error in the number of NS-NS mergers that form in GCs with a given
initial mass.}
\label{gc:mass-number}
\end{figure}

Each model was checked for the number of NS-NS mergers that occur inside 
the cluster and the number of escaping NS-NS binaries that would merge within a 
Hubble time. Table~\ref{table:gc-models} provides a summary of the initial models 
that contributed to forming coalescing NS-NS binaries along with the range of their 
final masses and number of NS-NS binaries they produced. From the limited number 
of simulations, we found $34$ coalescing NS-NS binaries. 21 of these are escaping 
NS-NS binaries that merge within a Hubble time and 13 merge inside the cluster. 
Although, there are not too many merging NS-NS binaries from the simulated 
cluster models, we find that more massive clusters produce more such binaries 
(see Fig.~\ref{gc:mass-number}). While there is a large number of single NSs in 
high binary fraction models with $\lambda=0.05$ for CE compared to models
with $\lambda=0.0$, however, there is no significant dependence of number of
merging NS-NS systems on CE parameters or metallicity. Our most massive models,
with zero NS natal kicks can produce 4 coalescing NS-NS binaries. The number of 
NS-NS binaries is correlated with the initial mass of the GC:

\begin{equation}
  N_{\rm nsns} = 0.001 \times (M_{\rm gc}/\msun)^{0.55}.
\label{eq.gcmass-number}  
\end{equation}

\begin{figure}   
\hspace*{-0.3cm}   
\includegraphics[width=9.2cm]{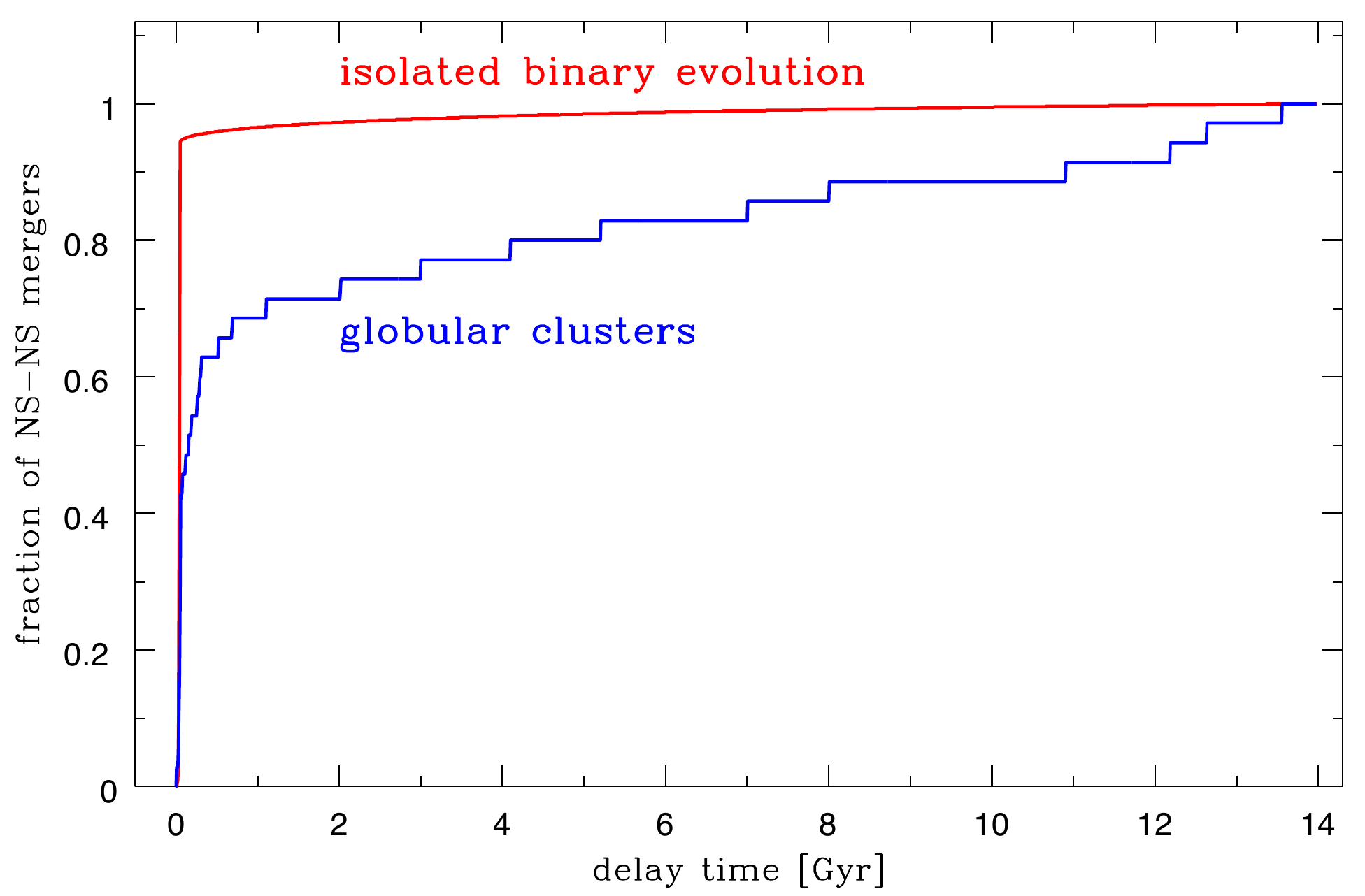}  
\caption{    
The normalized cumulative distribution of NS-NS merger delay time (from star 
formation to the merger) for classical isolated binary evolution and globular 
cluster formation channels. The isolated binary calculations formed 52,182 NS-NS 
mergers and there were 34 NS-NS mergers in GC simulations.}
\label{fig.tdel}
\end{figure}

The coalescence time distribution for these merging NS-NS binaries shows a peak 
within $1$ Gyr of cluster evolution. These are mostly NS-NS binaries that formed 
from the binary evolution of primordial binaries.
Fig.~\ref{fig.tdel} shows the cumulative distribution of merger 
times for NS-NS binaries originating in GCs. First $1$ Gyr of GC evolution 
produces $70\%$ of all our NS-NS mergers. Between $1$ Gyr and $11$ Gyr, we note $20\%$ 
of coalescing NS-NS binaries (nearly a uniform distribution in time).
Between $11$ Gyr to $14$ Gyr, a small peak is noted that contains $10\%$ of GC NS-NS
mergers. These late merging NS-NS binaries mostly form
because of dynamical interactions of NSs and other binary systems that begin to
segregate and form binaries in response to the core collapse of the GC as its
BH population starts to deplete. In many cases ($40\%$), the NSs in these coalescing
binary systems form from AIC of an oxygen-neon-magnesium WD or through mergers of WDs.
Formation of such a NS is illustrated in Figure~\ref{fig.nsns2}.
The small peak in merger rate between 11 to 13 Gyr is particularly interesting
as \citet{Blanchard2017} estimated using observations and stellar spectral population 
synthesis models that the star formation in NGC 4993 peaked 10 Gyr ago and that the 
median merger time for GW170817 may be as high as 11.2 Gyr.

For the rate calculation presented in Sec.~\ref{sec.gc} we made many approximations
to get a first order estimate of the merger rate. It is assumed that 
GCs comprise of $0.02\%$ of the current stellar mass in all elliptical galaxies 
within 100 Mpc. This assumption is based on the observed relations between the 
total stellar mass of an elliptical galaxy and the amount of mass in GCs 
\citep[see Fig. 8 in]{Harris2015}. We know from Milky Way GCs that their initial 
mass had to be larger to account for their current masses. Assuming that on 
average, GCs were up to 5 times more massive than their current mass \cite{Webb2015}, 
we take that $1\%$ of all initial stellar mass in ellipticals is in GCs. We further 
assume that the initial 
GC mass distribution follows the observed log-normal luminosity distribution of 
GCs that was observed by \citet{Harris2014}. This log-normal luminosity 
distribution was used by \citet{Rodriguez2016b,Rodriguez2015} to obtain a current 
log-normal mass function for GCs assuming a M/L ratio of 2. Taking a log-normal 
distribution with mean mass value of $\log_{10} (M) = 5.54$ and $\sigma_{M}=0.52$ 
\citet{Rodriguez2016b}, we sample GC masses and multiply each of them by a
factor of 5 to obtain a cumulative initial mass of $4.5 \times 10^{12} M_{\odot}$. 
Initial masses for GCs are then used to estimate the number of coalescing NS-NS 
binaries that could potentially originate from those systems using the power-law 
relation shown in Equation~\ref{eq.gcmass-number}. We assume that all
GCs with initial masses lower than $3\times 10^{5} M_{\odot}$ will produce at least
1 coalescing NS-NS binary. We use the total number of coalescing NS-NS binaries
produced by all GCs and using the merger time distribution inferred from the few
coalescing NS-NS binaries that emerged from the GC models, we can estimate the expected
number of mergers in different time intervals.

Taking that $\sim 70\%$ of the mergers occur within the first 1 Gyr, we estimate merger 
rates for NS-NS binaries originating from GCs in elliptical galaxies in 100 Mpc 
to be $\sim 2 \times 10^{-3}$ yr$^{-1}$. For the 5 Gyr, rate calculation, we take 
that $6\%$ of the total NS-NS mergers occur around this time ($4$--$7$ Gyr). We find this
corresponds to merger rate $\sim 2 \times 10^{-4}$ yr$^{-1}$. For 9 to 11 Gyr, we 
take that $5\%$ of the mergers occur within this interval, we compute the rate to be 
$\sim 1 \times 10^{-4}$ yr$^{-1}$. Like field calculations, the rates decrease with aging 
population. However, for GC between 11 to 13 Gyr, there is an increase in the 
number of merging NS-NS binaries. If we assume that $10\%$ of the coalescing 
binaries will merge between 11 to 13 Gyr, then the rate at this interval is 
$\sim 3 \times 10^{-4}$ yr$^{-1}$. 

The rates presented here are based on many favorable assumptions and are 
optimistic. While, it could be possible that the contribution from GCs could be 
an order of magnitude higher if they made up for a higher fraction of the total 
stellar mass in elliptical galaxies, the natal kicks used in our model are much 
lower than the typical kicks derived from observations of proper motions of 
pulsars \citep{Hobbs2005}. High natal kicks for NSs would make it more difficult 
to retain them in GCs and this will significantly reduce the expected rate. It 
is possible to form NSs via other channels in dense environments through 
dynamical interactions, however, in order to do this GCs must undergo core 
collapse. In models that retain a high number of BHs and NSs, a BH subsystem can 
provide energy to the system preventing core collapse. Only during the later 
evolution of such clusters, when BHs have depleted do NSs start to segregate. BHs can 
quickly deplete in dense models with short half-mass relaxation time. However, 
initially dense GCs can form an intermediate-mass BH \citep{Giersz2015}
which can then deplete the population of compact objects in the cluster. In
order to properly and thoroughly investigate the production of NS-NS binaries 
in GCs, a larger set of simulations covering a larger initial parameter space 
is necessary.

\begin{table*}
\caption{Initial parameters for globular cluster models. All GCs were 
initially non-segregated \citet{King1966} models with central concentration 
parameter ($\rm W_{0}$) value of 6.} 
\label{table:gc-models}      
\centering                                     
\resizebox{1.0\textwidth}{!}{ 
\begin{tabular}{|c|c|c|c|c|c|c|c|}
\hline
\textbf{\begin{tabular}[c]{@{}c@{}}Number of Objects\\ ($\times 10^{5}$)\end{tabular}} & \textbf{\begin{tabular}[c]{@{}c@{}}Initial Binary\\  Fraction\end{tabular}} & \textbf{\begin{tabular}[c]{@{}c@{}}Initial Mass\\ ($M_{\odot}$)\end{tabular}} & \textbf{\begin{tabular}[c]{@{}c@{}}Half-Mass Radius\\ (Tidal Radius) (pc)\end{tabular}} & \textbf{\begin{tabular}[c]{@{}c@{}}Metallicity\\ (Z)\end{tabular}} & \textbf{NS Kicks} & \textbf{\begin{tabular}[c]{@{}c@{}}12 Gyr Mass\\ ($M_{\odot}$)\end{tabular}} & \textbf{\begin{tabular}[c]{@{}c@{}}Number of\\ merging NS-NS binaries\\ /Number of GC Models\end{tabular}} \\ \hline
1 & 0.95 & $1.2 \times 10^{5}$ & 1.2(60), 2.4(60) & 0.01 & 0.0, 100.0 & $3.5-3.8 \times 10^{4}$ & 3/4 \\ \hline
1 & 0.1 & $6.3 \times 10^{4}$ & 1.2(60), 2.4(60) & 0.01 & 0.0,100.0 & $1.8-2.0 \times 10^{4}$ & 1/4 \\ \hline
5 & 0.95 & $5.9 \times 10^{5}$ & 1.2(60), 2.4(60) & \begin{tabular}[c]{@{}c@{}}0.01, 0.002\\ CE($\lambda = 0, 0.05$  $\alpha=1$)\end{tabular} & 0.0,100.0 & $2.4-2.5 \times 10^{5}$ & 9/6 \\ \hline
5 & 0.1 & $3.2 \times 10^{5}$ & 1.2(60), 2.4(60) & \begin{tabular}[c]{@{}c@{}}0.01\\ CE($\lambda = 0, 0.05$,$\alpha=1$)\end{tabular} & 100.0 & $1.3-1.5 \times 10^{5}$ & 3/4 \\ \hline
10 & 0.95 & $1.2 \times 10^{6}$ & 1.2(60) & \begin{tabular}[c]{@{}c@{}}0.01,0.002\\ CE($\lambda = 0, 0.05$,$\alpha=1$)\end{tabular} & 0,0, 100.0 & $5.3-5.4 \times 10^{5}$ & 15/7 \\ \hline
10 & 0.1 & $6.4 \times 10^{5}$ & 1.2(60) & \begin{tabular}[c]{@{}c@{}}0.01\\ CE($\lambda = 0, 0.05$,$\alpha=1$)\end{tabular} & 100.0 & $2.9-3.0 \times 10^{5}$ & 3/2 \\ \hline
\end{tabular}
}
\end{table*}

\subsection{Example of Calculations}
\label{sec.app2b}

An example of a NS-NS binary that forms and merges inside a GC at 11 Gyr is 
shown in Figure~\ref{fig.nsns2}. The GC model in which this merging NS-NS 
formed had initially $1 \times 10^{6}$ objects with binary fraction of $95\%$, 
metallicity $Z = 0.01$, initial half-mass and tidal radii were $1.2$ pc and 
$60$ pc, respectively. NS natal kicks were given by a Maxwellian distribution 
with $\sigma=100\kms$, and CE parameters were $\alpha=1$ and $\lambda=0.05$. 
The merging NSs originated from two separate initial binaries in this GC model. 

The first NS formed formed as an end product of the evolution of an $11\msun$ 
star (primary) that was in a wide binary system with a $4.95\msun$ companion. 
This binary had an initial semi-major axis of $4937 \rsun$ and an eccentricity of 
$0.17$. After $31$ Myr of evolution, the primary star became a NS in a core-collapse 
SN (see left side of Fig.~\ref{fig.nsns2}). This $\sim 1.3\msun$ NS received a 
natal kick and became a single star, but was still retained in the GC. For 
the next $\sim 9.5$ Gyr, this NS remained in the GC and did not undergo any 
strong dynamical interactions. At $\sim 9.6$ Gyr, the NS approached the GC 
center as the GC evolved towards core collapse. At this time the NS 
undergoes a strong interaction with a binary system comprising of two main 
sequence stars with masses of $0.4\msun$ and $0.7\msun$. During this 
binary-single interaction, the $0.4\msun$ MS star merged with the NS 
resulting in the formation of a $1.7\msun$ NS. The $0.7\msun$ MS star then 
became the binary companion of this heavy NS. 

At $10.45$ Gyr, this NS-MS binary interacted with an Oxygen-Neon-Magnesium 
(ONeMg) WD of mass $1.3\msun$. This heavy WD had formed from the evolution of 
a $6.9\msun$ star (see right side of Fig.~\ref{fig.nsns2}). During the 
binary-single interaction between the NS-MS binary and the WD, the $0.7\msun$ 
MS star was exchanged from the binary and the WD took its place resulting in 
the formation of a NS-WD binary. At $10.60$ Gyr, the NS-WD binary interacted 
with a low mass MS star of $0.12\msun$. During this interaction, the 
$0.12\msun$ MS star merged with the WD forming an AGB star. Now the NS is in 
a CE binary with an AGB star. This CE binary had an orbital separation of
$1252\rsun$ and eccentricity $e=0.97$. During the CE, mass was transferred 
from the AGB onto the NS increasing the mass of the NS from 1.7 $M_{\odot}$ 
to $1.9\msun$. The CE phase exposed again the  $1.3\msun$ ONeMg WD (the AGB star 
envelope successfully ejected). The post-CE NS-WD binary
circularized ($e=0$) during the CE phase and had an orbital separation of 
$3\rsun$. At $10.7$ Gyr, this NS-WD binary interacts with a Carbon-Oxygen
(CO) WD with mass $0.68\msun$. During this interaction, the ONeMg WD 
merges with the CO WD resulting in the formation of a NS of $1.26\msun$ 
due to accretion induced collapse (AIC). We assume no natal kick in the AIC
NS formation process. Following this interaction, we get a NS-NS binary 
comprising of $1.9\msun$ and $1.26\msun$ NSs with orbital separation of 
$3.5\rsun$ and (dynamical interaction induced) eccentricity of $e=0.75$. 
This NS-NS binary merges inside the GC at $\sim 11$ Gyr due to gravitational 
wave emission. 

Most of the GC NS-NS mergers have short delay times: $70\%$ of the mergers have
delay times $\lesssim 1$ Gyr (see Fig.~\ref{fig.tdel}). These
mergers host NSs from the evolution of massive stars that have formed via
regular (iron) core-collapse SNe. NS-NS binaries that merge at later times 
($10$--$14$ Gyr) form typically in the way shown by the example discussed in 
this section. In some cases, both the NSs in the binary form via AIC of WDs 
or through ECS. Like in our example, in certain cases the mass of one of the 
NSs is increased due to a prior merger with another star. The formation of a
NS from the merger of an ONeMg WD with a CO WD could possibly produce a radio 
transient~\citep{Moriya2016} or a short gamma-ray burst~\citep{Lyutikov2017}.
The number of such binaries is too low in our simulated models to make 
reliable comparisons of physical properties of GC NS-NS mergers with NS-NS 
mergers formed in isolated binary evolution. In future studies, we plan on 
increasing statistics of our GC models to deliver thorough comparison of GC 
and isolated binary evolution NS-NS mergers.

\section{Nuclear Cluster Dynamics}
\label{sec.nc}

\subsection{Overall Description}

We use a semi-analytic approach for modeling nuclear cluster (NC) formation 
in galactic nuclei~\citep{ArcaSedda2014}, coupled with results on NS-NS mergers 
in GCs achieved through a series of {\tt MOCCA} models. We consider two basic 
scenarios of NC formation: dry-merger model via GCs segregation and mergers into 
galactic centers~\citep{Tremaine1975,Capuzzo1993,Antonini2013} and in-situ model 
from gas deposits in galactic centers~\citep{King2003,Bekki2007,Nayakshin2009}. 
A way to disentangle the two processes is to examine observational NC-host galaxy 
connections~\citep{Cote2006,Graham2012,Turner2012}. 

Taking advantage of 
semi-analytic techniques, several authors have shown that the dry-merger 
scenario provides theoretical correlation laws in good agreement with 
observations~\citep{Antonini2013,ArcaSedda2014b,Gnedin2014}. Moreover, a number 
of studies provided detailed numerical modeling of NC formation through 
dry-merger mechanisms in galaxies mass range typical of dwarf galaxies 
~\citep{ArcaSedda2016b,ArcaSedda2017}, Milky Way--like galaxies
~\citep{Antonini2012,ArcaSedda2015,ArcaSedda2017c,Tsatsi2017}, and massive 
ellipticals~\citep{ArcaSedda2017b}. In particular, the dry-merger scenario 
provides an excellent explanation for the observational dearth of NCs in the 
galaxy mass range $M_g>10^{11}\msun$, which is observed when the expected
supermassive BH mass overtakes the NC mass~\citep{Neumayer2012,ArcaSedda2014b}. 
Indeed, it has been shown that above this galaxy mass threshold the supermassive 
BH tidal force is sufficient to disrupt the infalling clusters and prevent the NC 
formation~\citep{Antonini2013,ArcaSedda2014b,ArcaSedda2016,ArcaSedda2017b, 
Antonini2015}.

It has been shown that the relation connecting the NC mass, $M_{\rm NC}$,  
and the host galaxy velocity, $\sigma_g$, dispersion was similar to the 
$M_{\rm NC} - \sigma_g$ relation well known for supermassive BHs~\citep{Ferrarese2006}. 
However, later studies based on database revealed that the NC 
$M_{\rm NC} - \sigma_g$ relation is much shallower~\citep{Leigh2012,Scott2013,
Georgiev2016}, thus suggesting that the processes at play for NC and supermassive 
BH formation are likely unrelated, at least in part. The observed 
$M_{\rm NC} - \sigma_g$ relation represents an unique tool to disentangle the 
possible NC formation scenarios. Moreover, as pointed out by~\cite{Rossa2006}, 
NCs are characterized by a complex star formation history, being characterized 
by an old stellar population with ages $\sim 10$ Gyr and a younger population, 
with estimated ages below $100$ Myr. This feature is also observed in the Milky 
Way NC, possibly suggesting that several bursts of in-situ star formation occurred 
over its entire lifetime~\citep{Baumgardt2018}.

Although the dry-merger scenario is proven well at explaining the observed NC
scaling relations, it is quite difficult to explain NC complex star formation 
history, which is instead well motivated under the in-situ scenario. In fact, 
it is generally believed that both processes are at play during NC formation, 
although it is rather difficult to determine which one dominates.

In the dry merger model we assume that $f_{\rm a}=0.7$ of elliptical galaxies 
have NCs, that $f_{\rm b}=0.01$ of the total galaxy mass is found in GCs and 
that only some fraction ($f_{\rm c}$) of GCs contribute to the formation of NCs. 
This gives the total stellar mass in NCs found in all elliptical galaxies within 
$100$ Mpc: $M_{\rm nc,tot} = f_{\rm a} f_{\rm b} f_{\rm c} \ M_{\rm ell,tot}$. 
Each elliptical galaxy is populated with GCs with masses as given in 
Sec.~\ref{sec.gc} and the number of GCs per host galaxy is proportional to host 
mass. We examine which GCs have a dynamical friction timescale shorter than the 
tidal disruption timescale in a given host galaxy, which is the typical time over 
which galactic tidal forces drive the GC dissolution. Each such GC is assumed to 
contribute its mass (and NS-NS mergers given by {\tt MOCCA} simulations) to the 
host galaxy NC. We find wide range of NC masses: $M_{\rm nc}= 10^4$--$10^8\msun$
with a typical mass of $M_{\rm nc,ave}= 10^7\msun$.

\begin{figure}   
\hspace*{-0.3cm}   
\includegraphics[width=9.2cm]{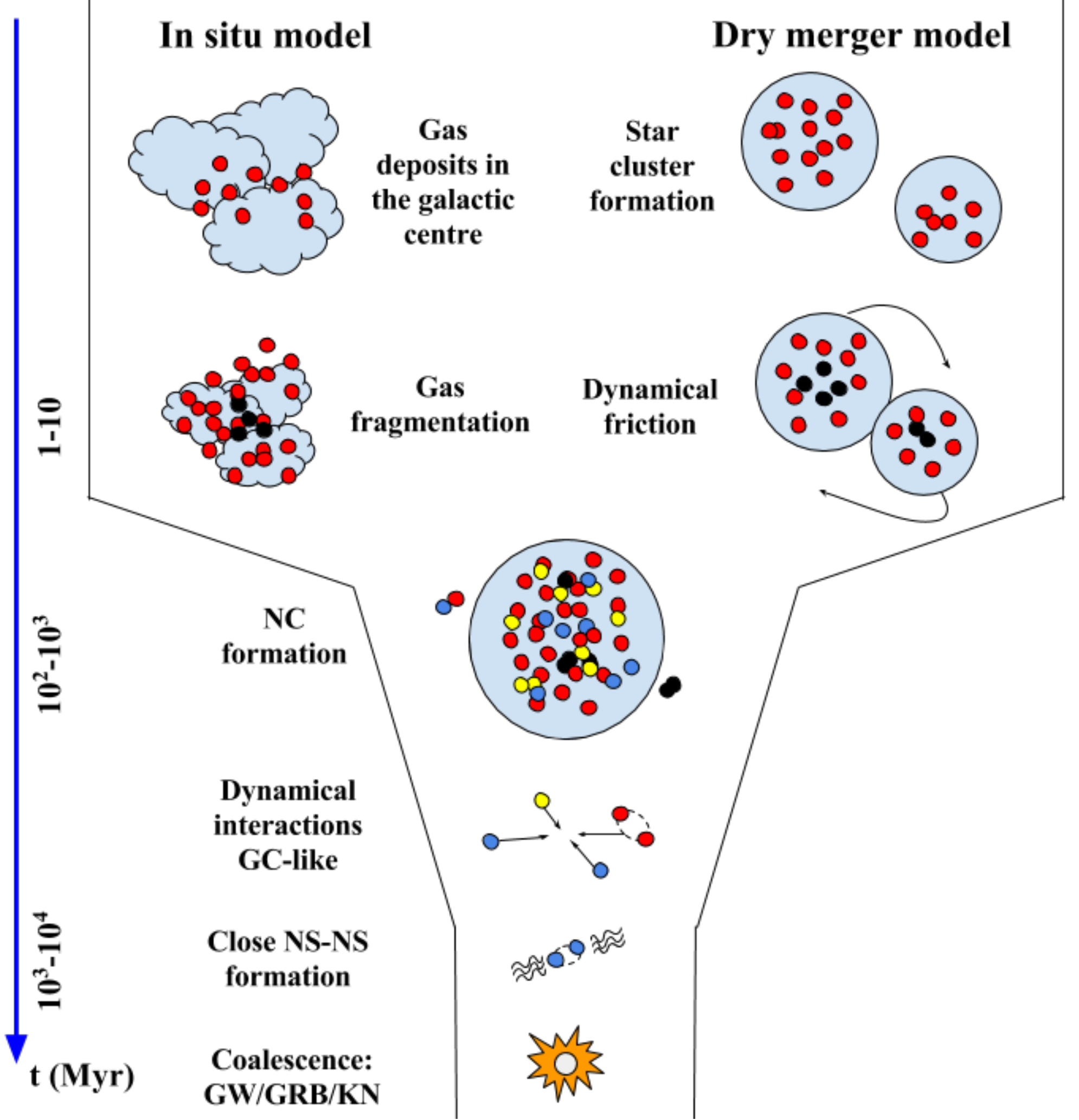}  
\caption{    
Example of the formation of a NS-NS merger similar to GW170817 in a 
nuclear cluster. 
}
\label{fig.nsns3}
\end{figure}

We assume that elliptical galaxy masses are distributed in the range $M_{\rm ell}= 
10^8$--$10^{12}\msun$ according to a Schechter function with typical parameters 
drawn according to observations of the local Universe~\citep{Conselice2006}. 
Varying the slope of the mass density profile and effective galaxy radius and 
averaging over the galaxy mass range, characterized by a mean mass of elliptical 
galaxy ($M_{\rm ell,ave}=1.6\times 10^{9}\msun$), we get $f_{\rm c}= 0.17$ 
(see Sec.~\ref{sec.app3}).

NC formation occurs on a typically longer timescale ($\sim 20$--$200$ Myr:
\cite{ArcaSedda2015}) than formation of NS in core collapse ($\sim 10$--$50$ Myr), 
but on a shorter timescale than NS-NS merger dynamical formation ($>1$ Gyr). 
Therefore, if NS was subject to a strong natal kick and if it was removed 
from GC it does not contribute to the calculations of NC NS-NS merger rates. 
However, dynamical formation of NS-NS mergers is enhanced by high NC mass: 
$f_{\rm dyn}=(M_{\rm nc,ave} / M_{\rm gc,ave})^{0.55}=(10^7/1.8\times10^6)^{0.55}=2.6$, 
where we have used average GC and NC mass in our simulations. 
The NC NS-NS merger rate is then:  
\begin{equation}
R_{\rm nsns,nc} = f_{\rm dyn} \frac{M_{\rm nc,tot}}{M_{\rm gc,tot}} R_{\rm nsns,gc} = 
f_{\rm dyn} f_{\rm a} f_{\rm c}  R_{\rm nsns,gc} = 0.31 R_{\rm nsns,gc}.
\label{eq.nc}
\end{equation}
NC NS-NS merger rate is very small: $6 \times 10^{-5}$ yr$^{-1}$ for $t_{\rm sf}=5$ 
Gyr and this is due to the fact that only a small fraction ($f_{\rm c}=0.17$) of 
GCs contribute to the formation of a typical NC.  

In the in-situ model, we assume that NC masses are the same as in the dry-merger 
model. The only boost to NS-NS merger rate in in-situ model comes then from higher 
retention fraction of NSs. The typical retention fraction of NSs formed in 
supernovae (subject to a potential natal kick) is $f_{\rm ns} \sim 0.3$
for our GC assumptions (natal kicks with $\sigma=100\kms$ for core-collapse
supernovae and $0\kms$ for electron capture supernovae). Note that a
fraction: $f_{\rm aic} \sim 0.5$  of NSs form from white dwarfs in GCs
without a natal kick (either in white dwarf mergers or during accretion induced 
collapse during mass transfer in close binary). Therefore, if we allow all NSs 
remain in NC the rate increase may be estimated as $(1-f_{\rm ns}) (1-f_{\rm aic}) 
=0.35$. It is expected that some NCs form via dry-mergers and some in-situ. Even 
if the majority of mergers from in-situ, the rate increase to the rate estimate 
given by Eq.~\ref{eq.nc} is negligible in the context of our study ($\lesssim 35\%$) 
and we neglect it in the values reported in Table~\ref{tab.rates}.

\subsection{Details of Calculations}
\label{sec.app3}

We made use of the semi-analytic approach described by \cite{ArcaSedda2014} to 
calculate the NC masses as a function of their host galaxy mass. To do this, we 
created 2750 galaxy models at varying galaxy total mass, inner slope of the 
density profile and galaxy effective radius.

In order to model the galaxy we used the \cite{Dehnen1993} family of 
potential-density pairs, whose density profile is given by:
\begin{equation}
\rho(r) = \frac{(3-\gamma)M_g}{4\pi r_g^3}\left(\frac{r}
{r_g}\right)^{-\gamma}\left(1+\frac{r}{r_g}\right)^{-4+\gamma},
\end{equation}
where $\gamma$ is the inner slope of the galaxy density profile, $M_g$ is the 
galaxy total mass and $r_g$ its length scale which is connected to the galaxy 
effective radius through the relation 
\begin{equation}
R_{\rm eff} =  \frac{3}{4}\frac{r_g}{2^{1/(3-\gamma)}-1}.
\end{equation}

For each galaxy model we selected $\gamma$ randomly between $0$ and $1$, in 
order to consider both cored and cuspy systems. The effective radius is 
varied according to the following relation 
\begin{equation}
R_e = A_g\left(\frac{M_g}{10^8\msun}\right)^{B_g},
\end{equation}
with $A_g = 0.706 \pm 0.005$ kpc and $B_g = 0.165 \pm 0.001$. This produces 
effective radii in quite good agreement with observed galaxies in terms of 
effective radii and velocity dispersions, as shown in \cite{ArcaSedda2014}.

Once the galaxy model has been set, we populate it with GCs, provided that 
the GC system total mass is 
\begin{equation}
\label{eq.gctot}
M_{\rm gc,tot} = 0.01 M_g.
\end{equation}
Our assumption relies upon the recent discussion arose by \cite{Webb2015}, 
which suggested that Galactic GCs were characterized at their birth by an 
initial mass at most $\sim 4.5$ times their current values, provided that 
their current $M_{\rm gc,tot}$ is $\lesssim 0.002 M_g$. 

The GC masses are kept according to a log-normal distribution with a 
mean value that depends on the galaxy mass through the relation 
\begin{eqnarray}
\label{eq.mgcave}
M_{\rm gc,ave} &=& 2.5\times 10^3 \msun \left(5-\log\frac{M_g} 
{10^8\msun}\right) \times \\
  & & \times \left[ 1 + 0.08 \left(\frac{M_g}{10^8\msun}\right)^{0.75} 
\left(8+\log\frac{M_g}{10^8\msun}\right) \right].\nonumber
\end{eqnarray}
This equation allows modeling the GC mass distribution while taking into 
account the fact that smaller galaxies host, on average, smaller GCs. This 
choice leads to $M_{\rm gc,ave}\sim 2.3\times 10^4\msun$ for galaxies with 
$M_g=10^8\msun$, and $M_{\rm gc,ave}\simeq 1.6-1.8\times 10^6\msun$ for galaxies 
in the mass range $10^{11}-10^{12}\msun$.

\begin{figure}
\centering
\includegraphics[width=9.2cm]{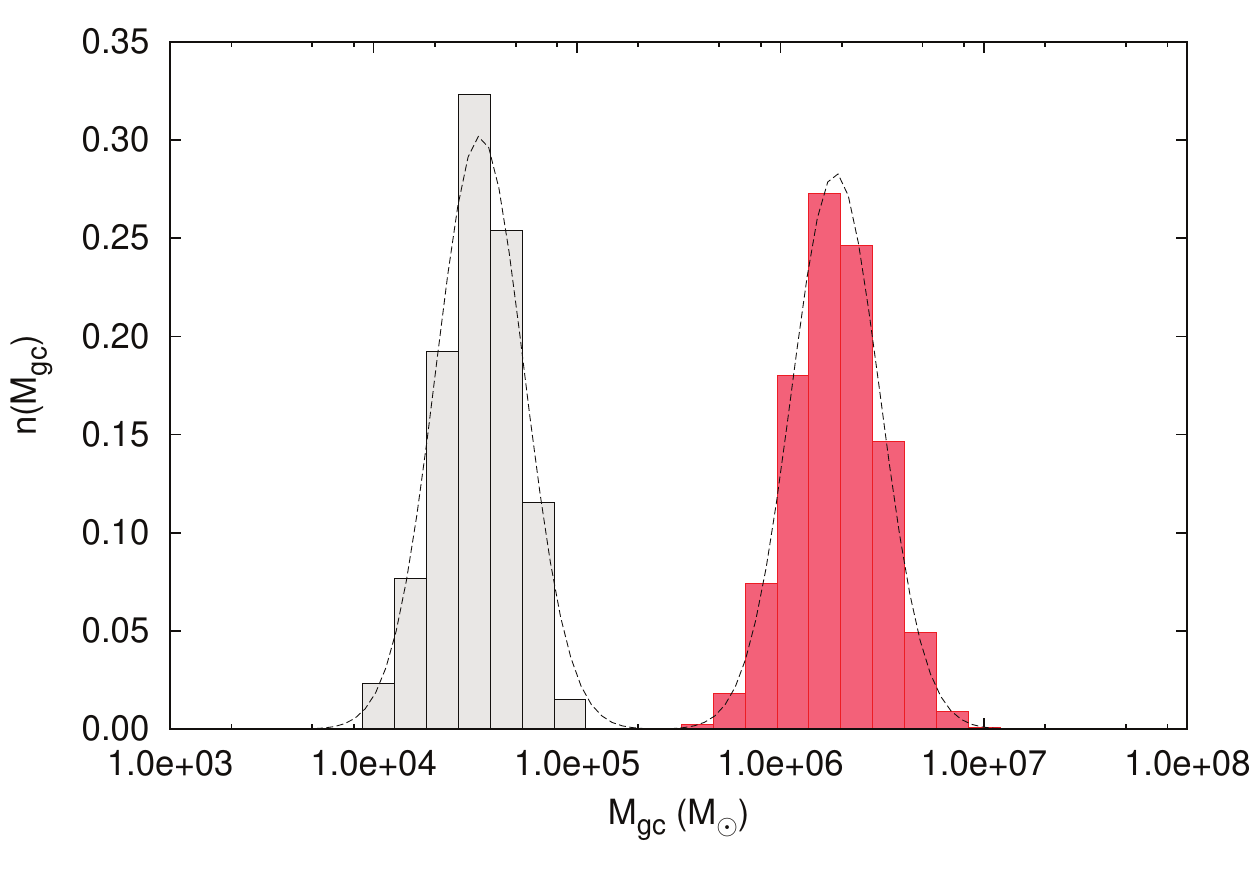}
\caption{GC mass distribution in one of the smallest 
($M_g\sim 3.2\times 10^8\msun$, grey boxes) and largest 
($M_g=5\times 10^{11}\msun$, red boxes) galaxy models.}
\label{gcdist}
\end{figure}

A further constrain that we required in the GC sampling process is that their 
mass density distribution follows the same mass density profile as this of the
host galaxy. In principle, there is no reason for assuming that the GC population 
formed with a different distribution compared to galactic stars. We note that our 
choice of GC system density profile leads to very good agreement with observed 
NCs~\citep{ArcaSedda2014}.

For each GC in our galaxy model, we calculated two typical time-scales: the 
dynamical friction time-scale $t_{\rm df}$, which represent the time over which 
the GC orbitally segregate to the galactic center, and the tidal disruption time 
$t_{\rm td}$, i.e. the time over which the galactic field drives the GC 
dissolution. The $t_{\rm df}$ has been calculated according to the approach described 
in \cite{ArcaSedda2014b}, which have shown through theoretical arguments and 
numerical simulations how $t_{\rm df}$ is connected to the galactic properties 
and the GC orbit:
\begin{equation}
t_{\rm df} = t_0 
g(e_{\rm gc},\gamma)
\left( \frac{ M_{\rm gc}}{M_g} \right)^{-0.67}
\left( \frac{ r_{\rm gc}}{r_g} \right)^{1.74},
\end{equation}
where $t_0$ is a normalization factor, $g(e_{\rm gc},\gamma)$ is a function 
connecting the eccentricity of the GC orbit in the galaxy and the galaxy 
density slope, and $r_{\rm gc}$ the GC orbital radius. This formula showed a 
remarkably good agreement with N-body simulations tailored to dwarf galaxies
~\citep{ArcaSedda2016,ArcaSedda2017d}, normal galaxies~\citep{ArcaSedda2015,
Petts2015,Petts2016,ArcaSedda2017c} and massive ellipticals~\citep{ArcaSedda2017b,
ArcaSedda2016b}. The tidal disruption time is calculated as the minimum between 
the two-body relaxation dissolution time~\citep{Lamers2010} and the time over 
which the GC dissolves due to repeated passage at pericenter within the host 
galaxy~\citep{ArcaSedda2014}.

Our approach leads to a clear correlation connecting the NC and the galaxy 
stellar masses, which is given by
\begin{equation}
\log \frac{M_{\rm nc}}{\msun} = A_{\rm nc}\log \frac{M_*}{\msun} 
+B_{\rm nc},
\end{equation}
with $A_{\rm nc}=1.000\pm0.005$ and $B_{\rm nc}=-3.17\pm0.05$, very similar to the 
most recent observational correlations~\citep{Scott2013,Georgiev2016,Capuzzo2017} 
and to earlier theoretical estimates~\citep{ArcaSedda2014, Gnedin2014}. 

\begin{figure}
\centering 
\includegraphics[width=9.2cm]{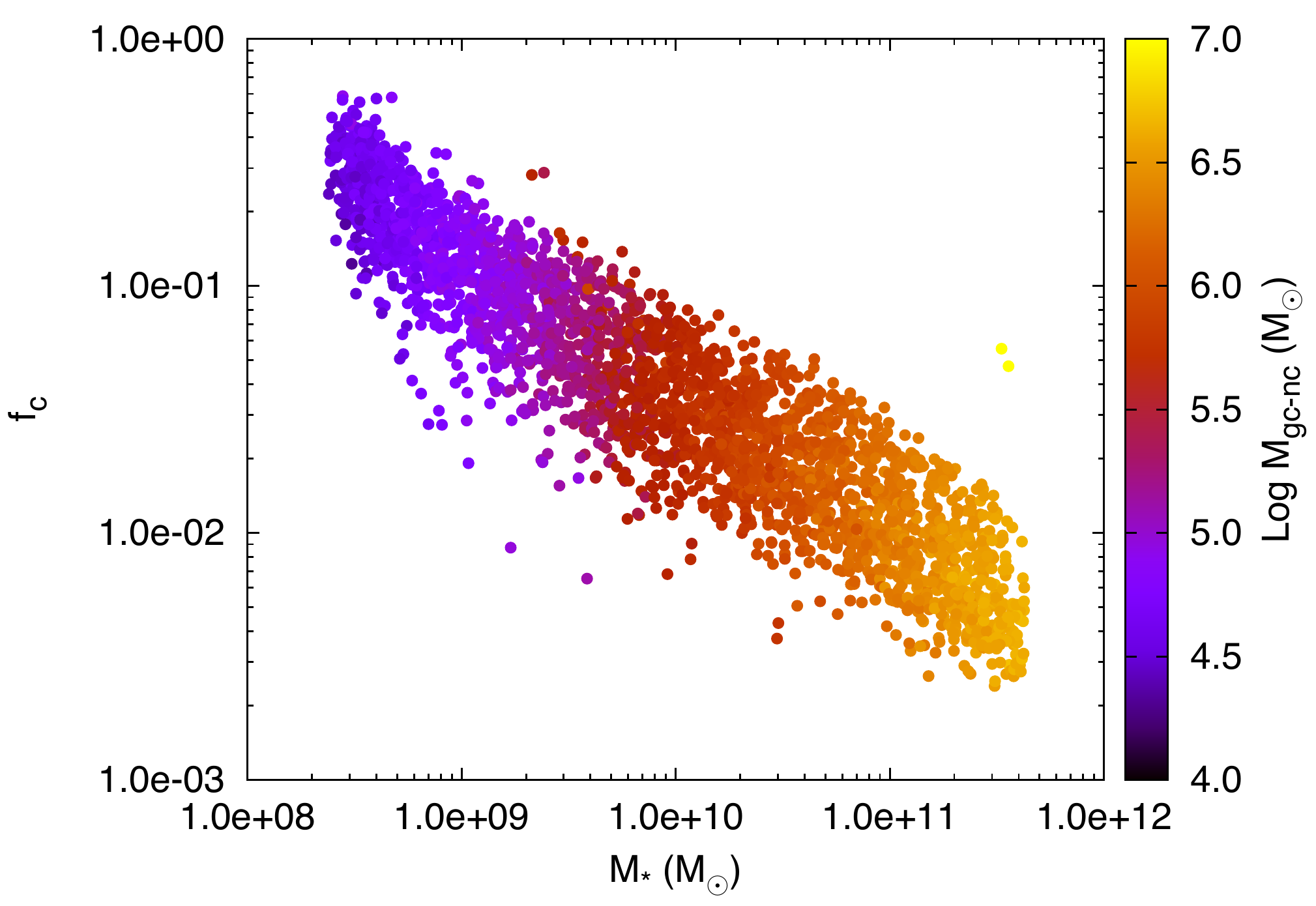}
\caption{NC mass normalized to the total GCs mass, as a function of the 
stellar galaxy mass. The color coded map shows the mean mass of GCs that 
contributed to the NC assembly.}
\label{corr}
\end{figure}

A crucial quantity needed to calculate the NS-NS merger rate for NCs is the 
fraction of the total GCs population mass that ends up in the NC, 
$f_{\rm c}(M_g) = M_{\rm nc}/M_{\rm gc,tot}$. Figure~\ref{corr} shows how this 
quantity varies at varying  galaxy stellar mass.
To convert our total galaxy masses into galaxy stellar masses we used the
relation~\citep{Gallazzi2006}
\begin{equation}
\log M_{*} = (0.783\pm 0.019) \log M_{g} + 2.19.
\label{gllz}
\end{equation}

As discussed above, our approach relies upon several specific assumptions that 
may affect the calculation of $f_{\rm c}$. In order to partly alleviate this 
issue, we assume a $30\%$ error in calculating $f_{\rm c}(M_{*})$, and note 
that its relation is enclosed within two power-laws
\begin{equation}
f_{\rm c}(M_*) = \alpha_{\rm max,min} M_*^\beta,
\end{equation}
with $\beta = -0.5$, $\alpha_{\rm max} = 1.7\times 10^{-2}$ and $\alpha_{\rm min} 
= 0.55\times 10^{-2}$. In the following, we make use of these two limiting 
quantities to calculate the average $f_{\rm c}(M_*)$ value in the local Universe.

Assuming that galaxies in the local Universe are distributed according to 
some mass function $\phi_g(M_*)$, we can average $f_{\rm c}(M_*)$ over the 
galaxy population:
\begin{equation}
\langle f_{\rm c} \rangle = \displaystyle{ 
\frac{\int_{M_{*1}}^{M_{*2}} f_{\rm c}(M_*)\phi_{\rm g}(M_*) dM_*}
{\int_{M_{*1}}^{M_{*2}}\phi_g(M_*) dM_*} ,
}
\end{equation}
with $M_{*1} = 10^8\msun$ and $M_{*2} = 10^{12}\msun$ being the minimum and 
maximum considered galaxy mass. A typical galaxy mass distribution is the 
\cite{Schechter1976} mass function, according to which 
$\phi_g(M_*)\propto (M_*/M_s)^{1+A}\exp(-M_*/M_s)$. 
We used the parameters provided by \cite{Conselice2016} for galaxies at 
redshift $z<0.2$: $A=-1.19$ and the mass scale $\log M_s = 11.20$. The 
solution of the Equation above is given by
\begin{equation}
\langle f_{{\rm c},i} \rangle = \displaystyle{ 
\alpha_i\frac
{\Gamma(1+A+\beta,\mu_{*2}) - \Gamma(1+A+\beta,\mu_{*1})}{\Gamma(1+A,\mu_{*2}) 
- \Gamma(1+A,\mu_{*1})}},
\end{equation}
where the subscript $i$ refers either to the maximum or minimum value of 
$\alpha$, $\Gamma(a,b)$ is the incomplete gamma function and 
$\mu_{*i} = M_{*i}/M_s$. Averaging over the maximum and minimum value for 
$\alpha$ we get 
\begin{equation}
\langle f_{\rm c} \rangle = 0.5\left(\langle f_{\rm c,{\rm max}} \rangle+\langle 
f_{\rm c,{\rm min}} \rangle\right)=0.17
\end{equation}
This implies that only $\sim 17\%$ of the whole GCs initial population 
contribute to the NC formation in the dry-merger model, making such channel
for NS-NS merger the least effective among the other proposed here. 
Note that our results depend on the choice of the initial GC mass function, 
which relies upon the presently observed GC distribution, in and out the
Galaxy. A wider mass function may possibly increase the number of GCs falling 
into the growing NC, leading $f_{\rm c}$ to increase. 
On the other hand, NCs are the densest stellar systems observed in the Universe, 
and this may prevent NS-NS formation via dynamical interactions, or at least 
cause delay, by the presence of a central dense BH subsystem or supermassive BH 
seed. 

\subsection{Example of Calculation}
\label{sec.app3b}

The NGC 4993 galaxy has a stellar mass $M_{*} = 4.4\times 10^{10} \msun$
~\citep{Blanchard2017}; the corresponding dynamical mass can be calculated 
using Equation \ref{gllz}.
Hence, according to Eq.~\ref{eq.gctot} the initial mass in GC will be
\begin{equation}
M_{\rm gc,tot} = 0.01 M_g = 3.3\times 10^8 {\rm M}_\odot,
\end{equation}
while the GC initial average mass is given by Eq.~\ref{eq.mgcave} 
\begin{equation}
M_{\rm gc,ave} = 4.1\times 10^5 {\rm M}_\odot,
\end{equation}
thus implying a number of GCs ending up in the galaxy nuclear star cluster 
\begin{equation}
N_{\rm gc,NSC} = f_c M_{\rm gc,tot} / M_{\rm gc,ave} \simeq 89
\end{equation}

Using Eq.~\ref{eq.gcmass-number} we can calculate the total number of 
potential NS-NS mergers in this galaxy as
\begin{equation}
N_{nsns} = \left(0.001M_{\rm gc,ave}^{0.55}\right)N_{\rm gc,NSC} = 108.
\end{equation}

As discussed in Sec.~\ref{sec.app3}, $\sim 70\%$ of these 108 NS-NS binaries 
will merge in the first 1 Gyr of the nuclear star cluster lifetime, $\sim 20\%$ in 
the time interval 1-11 Gyr and $\sim 10\%$ in the time range 11-14 Gyr.

\section{Discussion and Conclusion}
\label{sec.dis}

Theoretically obtained NS-NS merger rates may be compared with several 
empirically based estimates. For example, our earlier analysis of the isolated 
binary evolution channel shows perfect agreement with the observed Milky Way 
population of NS-NS systems, and acceptable agreement with short Gamma-ray 
burst rate estimates~\citep{Chruslinska2018}\footnote{However note that 
the optimistic model adopted here tends to overestimate rates of Galactic 
NS-NS systems, while being consistent with short Gamma-ray burst rates; see 
\cite{Chruslinska2018} Fig.6 and Fig.7: model J5 submodel B.}. Inferences on the 
NS-NS merger rate can be also made from measurements of metal enrichment in local 
Universe~\citep{Cote2018}. All these empirically based estimates are subject 
to large uncertainties. Pulsar beaming and luminosity function limit estimates 
of Galactic NS-NS merger rates, beaming and luminosity function and an unknown
contribution of BH-NS mergers limit short Gamma-ray burst NS-NS merger rates, 
and merger ejection mass along with an unknown contribution of supernovae
limit the inferences from r-process element observations. In this study we 
limit our {\em Discussion} to comparison of theoretically estimated NS-NS 
merger rates with gravitational wave data only. This datum (the LIGO/Virgo single 
detection) is currently the only direct measure of NS-NS merger rate.

Our numerical simulations indicate that the formation of NS-NS mergers in old
stellar populations, although possible, is unlikely to recover the merger rate 
inferred from the detection of GW170817. This is surprising as the three tested 
NS-NS formation mechanisms: classical isolated binary evolution, dynamical 
evolution in globular clusters, and nuclear cluster formation scenarios can 
reproduce the gravitational wave estimate of BH-BH merger rate. 

It is noted that NGC 4993 shows some shell structures and dust lanes that may be 
possibly indicative of a recent merger ($200$--$400$ Myr ago) with another smaller 
galaxy~\citep{Palmese2017,Ebrova2018}. If there was a recent burst of massive star 
formation in NGC 4993 induced by galaxy merger, it is possible that GW170817 was 
formed in such an event and our analysis and conclusions do not hold. Our results 
and the rest of our discussion are based on the assumption that late type galaxies 
do not experience recent vigorous ($\lesssim 1$ Gyr) star formation.

It can not be excluded that GW170817 is a BH-NS merger as the primary compact 
object ($1.36$--$2.26\msun$: LIGO/Virgo $90\%$ credible limit; ~\cite{GW170817}) 
may be a BH. Detailed examination of BH-NS formation models is desired for all 
three mechanisms. However, the existing models do not indicate that changing the 
identity of GW170817 could solve the tension. For example, in the classical 
binary evolution the local ($z \approx 0$) BH-NS merger rate density is smaller 
(or at best comparable) to the NS-NS rates~\citep{Belczynski2017b}. The rest of 
the discussion is based on the assumption that GW170817 is a NS-NS merger. 

It is possible that the LIGO/Virgo detection of GW170817 is not a statistical 
coincidence, but that finding the first NS-NS merger in an old host galaxy is.
In such a case isolated classical binary evolution can {\em marginally} explain 
the LIGO/Virgo observation. Population synthesis results show that if the {\em
entire} star formation (in old and young galaxies combined) is considered, then
theoretical rates may reach as high as $600\gpy$~\citep{Chruslinska2018} or 
$400\gpy$~\citep{Kruckow2018} and thus are consistent with LIGO/Virgo lower
$90\%$ credible limit of $320\gpy$. If this is the case, then future LIGO/Virgo 
detections will show prevalence of NS-NS merger detections associated with host
galaxies with ongoing (or recent) star formation. The rest of the discussion
is based on the assumption that association of GW170817 with old host galaxy
is not a statistical coincidence.

It cannot be excluded that the actual NS-NS merger rate is outside of the 
LIGO/Virgo $90\%$ credible limit ($1540^{+3200}_{-1220} \gpy$), but the
detection was made nevertheless (i.e., detection itself is a statistical
coincidence). In this case our models indicate that GW170817 was most likely 
formed in old host galaxy through the classical isolated binary evolution, 
that offers ($\gtrsim 100$ times) higher NS-NS merger rate than the dynamical 
formation scenarios in globular and nuclear clusters. The observational run O3 
(2018/19) should clarify this open issue as the increased LIGO/Virgo sensitivity 
and new NS-NS merger detections (or lack thereof) will place a better constraint 
on the NS-NS merger rate. 

However, if the NS-NS merger rate turns out to be as high as the most likely 
value of the LIGO/Virgo estimate ($1540\gpy$) it will indicate that our current 
understanding of formation process of NS-NS mergers in the three considered 
scenarios is not complete. Either the initial properties of binaries were different 
in the past when stars were forming in NGC 4993, or the evolutionary processes that 
lead to the NS-NS merger formation are not yet understood, or possibly a solution 
exists within current input physics and associated uncertainties, but was not yet 
found within the multi-dimensional parameter space. All these possibilities will need 
to be assessed and tested to inform our concepts of physics as derived from 
gravitational wave observations.

If all the above fails, other non-standard NS-NS merger formation scenarios must 
be considered and developed.

\begin{acknowledgements}
We thank anonymous referee for detailed and constructive comments. 
We would like to thank thousands of {\tt Universe@home} users that have provided 
their personal computers and phones for our simulations, and in particular to K. 
Piszczek (program IT manager). 
Authors acknowledge support from the Polish National Science Center (NCN) grants: 
Sonata Bis 2 DEC-2012/07/E/ST9/01360 (KB, MC, GW), Harmonia 6 UMO-2014/14/M/ST9/00707 
(KB),  Opus UMO-2016/23/B/ST9/02732 (AA, MG), UMO-2015/17/N/ST9/02573 (AA), and 
German Research Foundation (DFG) through the Sonderforschungsbereich SFB 881 grant 
(MAS). GW is partly supported by the President's International Fellowship Initiative 
(PIFI) of the Chinese Academy of Sciences under grant no.2018PM0017 and by the 
Strategic Priority Research Program of the Chinese Academy of Science “Multi-waveband 
Gravitational Wave Universe” (Grant No. XDB23040000).
\end{acknowledgements}

\bibliographystyle{aa}
%\bibliography{biblio}

\begin{thebibliography}{134}
\expandafter\ifx\csname natexlab\endcsname\relax\def\natexlab#1{#1}\fi

\bibitem[{Abbott {et~al.}(2017)Abbott, Abbott, Abbott, Acernese, Ackley, Adams,
  Adams, Addesso, Adhikari, Adya, Affeldt, Afrough, Agarwal, Agathos, Agatsuma,
  Aggarwal, Aguiar, Aiello, Ain, Ajith, Allen, Allen, Allocca, Altin, Amato,
  Ananyeva, Anderson, Anderson, Angelova, Antier, Appert, Arai, Araya, Areeda,
  Arnaud, Arun, Ascenzi, Ashton, Ast, Aston, Astone, Atallah, Aufmuth, Aulbert,
  AultONeal, Austin, Avila-Alvarez, Babak, Bacon, Bader, Bae, Bailes, Baker,
  Baldaccini, Ballardin, Ballmer, Banagiri, Barayoga, Barclay, Barish, Barker,
  Barkett, Barone, Barr, Barsotti, Barsuglia, Barta, Barthelmy, Bartlett,
  Bartos, Bassiri, Basti, Batch, Bawaj, Bayley, Bazzan, B\'ecsy, Beer, Bejger,
  Belahcene, Bell, Berger, Bergmann, Bernuzzi, Bero, Berry, Bersanetti,
  Bertolini, Betzwieser, Bhagwat, Bhandare, Bilenko, Billingsley, Billman,
  Birch, Birney, Birnholtz, Biscans, Biscoveanu, Bisht, Bitossi, Biwer,
  Bizouard, Blackburn, Blackman, Blair, Blair, Blair, Bloemen, Bock, Bode,
  Boer, Bogaert, Bohe, Bondu, Bonilla, Bonnand, Boom, Bork, Boschi, Bose,
  Bossie, Bouffanais, Bozzi, Bradaschia, Brady, Branchesi, Brau, Briant,
  Brillet, Brinkmann, Brisson, Brockill, Broida, Brooks, Brown, Brown, Brunett,
  Buchanan, Buikema, Bulik, Bulten, Buonanno, Buskulic, Buy, Byer, Cabero,
  Cadonati, Cagnoli, Cahillane, Calder\'on~Bustillo, Callister, Calloni, Camp,
  Canepa, Canizares, Cannon, Cao, Cao, Capano, Capocasa, Carbognani, Caride,
  Carney, Carullo, Casanueva~Diaz, Casentini, Caudill, Cavagli\`a, Cavalier,
  Cavalieri, Cella, Cepeda, Cerd\'a-Dur\'an, Cerretani, Cesarini, Chamberlin,
  Chan, Chao, Charlton, Chase, Chassande-Mottin, Chatterjee, Chatziioannou,
  Cheeseboro, Chen, Chen, Chen, Cheng, Chia, Chincarini, Chiummo, Chmiel, Cho,
  Cho, Chow, Christensen, Chu, Chua, Chua, Chung, Chung, Ciani, Ciolfi,
  Cirelli, Cirone, Clara, Clark, Clearwater, Cleva, Cocchieri, Coccia, Cohadon,
  Cohen, Colla, Collette, Cominsky, Constancio, Conti, Cooper, Corban, Corbitt,
  Cordero-Carri\'on, Corley, Cornish, Corsi, Cortese, Costa, Coughlin,
  Coughlin, Coulon, Countryman, Couvares, Covas, Cowan, Coward, Cowart, Coyne,
  Coyne, Creighton, Creighton, Cripe, Crowder, Cullen, Cumming, Cunningham,
  Cuoco, Dal~Canton, D\'alya, Danilishin, D'Antonio, Danzmann, Dasgupta,
  Da~Silva~Costa, Dattilo, Dave, Davier, Davis, Daw, Day, De, DeBra, Degallaix,
  De~Laurentis, Del\'eglise, Del~Pozzo, Demos, Denker, Dent, De~Pietri,
  Dergachev, De~Rosa, DeRosa, De~Rossi, DeSalvo, de~Varona, Devenson,
  Dhurandhar, D\'{\i}az, Dietrich, Di~Fiore, Di~Giovanni, Di~Girolamo,
  Di~Lieto, Di~Pace, Di~Palma, Di~Renzo, Doctor, Dolique, Donovan, Dooley,
  Doravari, Dorrington, Douglas, Dovale~\'Alvarez, Downes, Drago,
  Dreissigacker, Driggers, Du, Ducrot, Dudi, Dupej, Dwyer, Edo, Edwards,
  Effler, Eggenstein, Ehrens, Eichholz, Eikenberry, Eisenstein, Essick,
  Estevez, Etienne, Etzel, Evans, Evans, Factourovich, Fafone, Fair, Fairhurst,
  Fan, Farinon, Farr, Farr, Fauchon-Jones, Favata, Fays, Fee, Fehrmann, Feicht,
  Fejer, Fernandez-Galiana, Ferrante, Ferreira, Ferrini, Fidecaro, Finstad,
  Fiori, Fiorucci, Fishbach, Fisher, Fitz-Axen, Flaminio, Fletcher, Fong, Font,
  Forsyth, Forsyth, Fournier, Frasca, Frasconi, Frei, Freise, Frey, Frey,
  Fries, Fritschel, Frolov, Fulda, Fyffe, Gabbard, Gadre, Gaebel, Gair,
  Gammaitoni, Ganija, Gaonkar, Garcia-Quiros, Garufi, Gateley, Gaudio, Gaur,
  Gayathri, Gehrels, Gemme, Genin, Gennai, George, George, Gergely, Germain,
  Ghonge, Ghosh, Ghosh, Ghosh, Giaime, Giardina, Giazotto, Gill, Glover, Goetz,
  Goetz, Gomes, Goncharov, Gonz\'alez, Gonzalez~Castro, Gopakumar, Gorodetsky,
  Gossan, Gosselin, Gouaty, Grado, Graef, Granata, Grant, Gras, Gray, Greco,
  Green, Gretarsson, Groot, Grote, Grunewald, Gruning, Guidi, Guo, Gupta,
  Gupta, Gushwa, Gustafson, Gustafson, Halim, Hall, Hall, Hamilton, Hammond,
  Haney, Hanke, Hanks, Hanna, Hannam, Hannuksela, Hanson, Hardwick, Harms,
  Harry, Harry, Hart, Haster, Haughian, Healy, Heidmann, Heintze, Heitmann,
  Hello, Hemming, Hendry, Heng, Hennig, Heptonstall, Heurs, Hild, Hinderer, Ho,
  Hoak, Hofman, Holt, Holz, Hopkins, Horst, Hough, Houston, Howell, Hreibi, Hu,
  Huerta, Huet, Hughey, Husa, Huttner, Huynh-Dinh, Indik, Inta, Intini, Isa,
  Isac, Isi, Iyer, Izumi, Jacqmin, Jani, Jaranowski, Jawahar,
  Jim\'enez-Forteza, Johnson, Johnson-McDaniel, Jones, Jones, Jonker, Ju,
  Junker, Kalaghatgi, Kalogera, Kamai, Kandhasamy, Kang, Kanner, Kapadia,
  Karki, Karvinen, Kasprzack, Kastaun, Katolik, Katsavounidis, Katzman, Kaufer,
  Kawabe, K\'ef\'elian, Keitel, Kemball, Kennedy, Kent, Key, Khalili, Khan,
  Khan, Khan, Khazanov, Kijbunchoo, Kim, Kim, Kim, Kim, Kim, Kim, Kimbrell,
  King, King, Kinley-Hanlon, Kirchhoff, Kissel, Kleybolte, Klimenko, Knowles,
  Koch, Koehlenbeck, Koley, Kondrashov, Kontos, Korobko, Korth, Kowalska,
  Kozak, Kr\"amer, Kringel, Krishnan, Kr\'olak, Kuehn, Kumar, Kumar, Kumar,
  Kuo, Kutynia, Kwang, Lackey, Lai, Landry, Lang, Lange, Lantz, Lanza, Larson,
  Lartaux-Vollard, Lasky, Laxen, Lazzarini, Lazzaro, Leaci, Leavey, Lee, Lee,
  Lee, Lee, Lee, Lehmann, Lenon, Leon, Leonardi, Leroy, Letendre, Levin, Li,
  Linker, Littenberg, Liu, Liu, Lo, Lockerbie, London, Lord, Lorenzini,
  Loriette, Lormand, Losurdo, Lough, Lousto, Lovelace, L\"uck, Lumaca,
  Lundgren, Lynch, Ma, Macas, Macfoy, Machenschalk, MacInnis, Macleod, Maga\~na
  Hernandez, Maga\~na Sandoval, Maga\~na Zertuche, Magee, Majorana, Maksimovic,
  Man, Mandic, Mangano, Mansell, Manske, Mantovani, Marchesoni, Marion,
  M\'arka, M\'arka, Markakis, Markosyan, Markowitz, Maros, Marquina, Marsh,
  Martelli, Martellini, Martin, Martin, Martynov, Marx, Mason, Massera,
  Masserot, Massinger, Masso-Reid, Mastrogiovanni, Matas, Matichard, Matone,
  Mavalvala, Mazumder, McCarthy, McClelland, McCormick, McCuller, McGuire,
  McIntyre, McIver, McManus, McNeill, McRae, McWilliams, Meacher, Meadors,
  Mehmet, Meidam, Mejuto-Villa, Melatos, Mendell, Mercer, Merilh, Merzougui,
  Meshkov, Messenger, Messick, Metzdorff, Meyers, Miao, Michel, Middleton,
  Mikhailov, Milano, Miller, Miller, Miller, Millhouse, Milovich-Goff,
  Minazzoli, Minenkov, Ming, Mishra, Mitra, Mitrofanov, Mitselmakher,
  Mittleman, Moffa, Moggi, Mogushi, Mohan, Mohapatra, Molina, Montani, Moore,
  Moraru, Moreno, Morisaki, Morriss, Mours, Mow-Lowry, Mueller, Muir,
  Mukherjee, Mukherjee, Mukherjee, Mukund, Mullavey, Munch, Mu\~niz, Muratore,
  Murray, Nagar, Napier, Nardecchia, Naticchioni, Nayak, Neilson, Nelemans,
  Nelson, Nery, Neunzert, Nevin, Newport, Newton, Ng, Nguyen, Nguyen, Nichols,
  Nielsen, Nissanke, Nitz, Noack, Nocera, Nolting, North, Nuttall, Oberling,
  O'Dea, Ogin, Oh, Oh, Ohme, Okada, Oliver, Oppermann, Oram, O'Reilly,
  Ormiston, Ortega, O'Shaughnessy, Ossokine, Ottaway, Overmier, Owen, Pace,
  Page, Page, Pai, Pai, Palamos, Palashov, Palomba, Pal-Singh, Pan, Pan, Pang,
  Pang, Pankow, Pannarale, Pant, Paoletti, Paoli, Papa, Parida, Parker,
  Pascucci, Pasqualetti, Passaquieti, Passuello, Patil, Patricelli, Pearlstone,
  Pedraza, Pedurand, Pekowsky, Pele, Penn, Perez, Perreca, Perri, Pfeiffer,
  Phelps, Piccinni, Pichot, Piergiovanni, Pierro, Pillant, Pinard, Pinto,
  Pirello, Pitkin, Poe, Poggiani, Popolizio, Porter, Post, Powell, Prasad,
  Pratt, Pratten, Predoi, Prestegard, Prijatelj, Principe, Privitera, Prix,
  Prodi, Prokhorov, Puncken, Punturo, Puppo, P\"urrer, Qi, Quetschke, Quintero,
  Quitzow-James, Raab, Rabeling, Radkins, Raffai, Raja, Rajan, Rajbhandari,
  Rakhmanov, Ramirez, Ramos-Buades, Rapagnani, Raymond, Razzano, Read,
  Regimbau, Rei, Reid, Reitze, Ren, Reyes, Ricci, Ricker, Rieger, Riles, Rizzo,
  Robertson, Robie, Robinet, Rocchi, Rolland, Rollins, Roma, Romano, Romano,
  Romel, Romie, Rosi\ifmmode~\acute{n}\else \'{n}\fi{}ska, Ross, Rowan,
  R\"udiger, Ruggi, Rutins, Ryan, Sachdev, Sadecki, Sadeghian, Sakellariadou,
  Salconi, Saleem, Salemi, Samajdar, Sammut, Sampson, Sanchez, Sanchez,
  Sanchis-Gual, Sandberg, Sanders, Sassolas, Sathyaprakash, Saulson, Sauter,
  Savage, Sawadsky, Schale, Scheel, Scheuer, Schmidt, Schmidt, Schnabel,
  Schofield, Sch\"onbeck, Schreiber, Schuette, Schulte, Schutz, Schwalbe,
  Scott, Scott, Seidel, Sellers, Sengupta, Sentenac, Sequino, Sergeev,
  Shaddock, Shaffer, Shah, Shahriar, Shaner, Shao, Shapiro, Shawhan, Sheperd,
  Shoemaker, Shoemaker, Siellez, Siemens, Sieniawska, Sigg, Silva, Singer,
  Singh, Singhal, Sintes, Slagmolen, Smith, Smith, Smith, Somala, Son,
  Sonnenberg, Sorazu, Sorrentino, Souradeep, Spencer, Srivastava, Staats,
  Staley, Steinke, Steinlechner, Steinlechner, Steinmeyer, Stevenson, Stone,
  Stops, Strain, Stratta, Strigin, Strunk, Sturani, Stuver, Summerscales, Sun,
  Sunil, Suresh, Sutton, Swinkels, Szczepa\ifmmode~\acute{n}\else
  \'{n}\fi{}czyk, Tacca, Tait, Talbot, Talukder, Tanner, T\'apai, Taracchini,
  Tasson, Taylor, Taylor, Tewari, Theeg, Thies, Thomas, Thomas, Thomas, Thorne,
  Thorne, Thrane, Tiwari, Tiwari, Tokmakov, Toland, Tonelli, Tornasi,
  Torres-Forn\'e, Torrie, T\"oyr\"a, Travasso, Traylor, Trinastic, Tringali,
  Trozzo, Tsang, Tse, Tso, Tsukada, Tsuna, Tuyenbayev, Ueno, Ugolini,
  Unnikrishnan, Urban, Usman, Vahlbruch, Vajente, Valdes, Vallisneri, van
  Bakel, van Beuzekom, van~den Brand, Van Den~Broeck, Vander-Hyde, van~der
  Schaaf, van Heijningen, van Veggel, Vardaro, Varma, Vass, Vas\'uth, Vecchio,
  Vedovato, Veitch, Veitch, Venkateswara, Venugopalan, Verkindt, Vetrano,
  Vicer\'e, Viets, Vinciguerra, Vine, Vinet, Vitale, Vo, Vocca, Vorvick,
  Vyatchanin, Wade, Wade, Wade, Walet, Walker, Wallace, Walsh, Wang, Wang,
  Wang, Wang, Wang, Ward, Warner, Was, Watchi, Weaver, Wei, Weinert, Weinstein,
  Weiss, Wen, Wessel, We\ss{}els, Westerweck, Westphal, Wette, Whelan,
  Whitcomb, Whiting, Whittle, Wilken, Williams, Williams, Williamson, Willis,
  Willke, Wimmer, Winkler, Wipf, Wittel, Woan, Woehler, Wofford, Wong, Worden,
  Wright, Wu, Wysocki, Xiao, Yamamoto, Yancey, Yang, Yap, Yazback, Yu, Yu,
  Yvert, Zadro\ifmmode~\dot{z}\else \.{z}\fi{}ny, Zanolin, Zelenova, Zendri,
  Zevin, Zhang, Zhang, Zhang, Zhang, Zhao, Zhou, Zhou, Zhu, Zhu, Zimmerman,
  Zucker, \& Zweizig}]{GW170817}
Abbott, B.~P., Abbott, R., Abbott, T.~D., {et~al.} 2017, Phys. Rev. Lett., 119,
  161101

\bibitem[{{Antonini}(2013)}]{Antonini2013}
{Antonini}, F. 2013, \apj, 763, 62

\bibitem[{{Antonini} {et~al.}(2015){Antonini}, {Barausse}, \&
  {Silk}}]{Antonini2015}
{Antonini}, F., {Barausse}, E., \& {Silk}, J. 2015, \apj, 812, 72

\bibitem[{{Antonini} \& {Perets}(2012)}]{Antonini2012}
{Antonini}, F. \& {Perets}, H.~B. 2012, \apj, 757, 27

\bibitem[{{Antonini} \& {Rasio}(2016)}]{Antonini2016}
{Antonini}, F. \& {Rasio}, F.~A. 2016, \apj, 831, 187

\bibitem[{{Arca Sedda} {et~al.}(2018){Arca Sedda}, {Askar}, \&
  {Giersz}}]{ArcaSedda2018a}
{Arca Sedda}, M., {Askar}, A., \& {Giersz}, M. 2018, ArXiv e-prints
  [\eprint[arXiv]{1801.00795}]

\bibitem[{{Arca-Sedda} \&
  {Capuzzo-Dolcetta}(2014{\natexlab{a}})}]{ArcaSedda2014b}
{Arca-Sedda}, M. \& {Capuzzo-Dolcetta}, R. 2014{\natexlab{a}}, \apj, 785, 51

\bibitem[{{Arca-Sedda} \&
  {Capuzzo-Dolcetta}(2014{\natexlab{b}})}]{ArcaSedda2014}
{Arca-Sedda}, M. \& {Capuzzo-Dolcetta}, R. 2014{\natexlab{b}}, \mnras, 444,
  3738

\bibitem[{{Arca-Sedda} \& {Capuzzo-Dolcetta}(2016)}]{ArcaSedda2016b}
{Arca-Sedda}, M. \& {Capuzzo-Dolcetta}, R. 2016, \mnras, 461, 4335

\bibitem[{{Arca-Sedda} \&
  {Capuzzo-Dolcetta}(2017{\natexlab{a}})}]{ArcaSedda2017d}
{Arca-Sedda}, M. \& {Capuzzo-Dolcetta}, R. 2017{\natexlab{a}}, \mnras, 464,
  3060

\bibitem[{{Arca-Sedda} \&
  {Capuzzo-Dolcetta}(2017{\natexlab{b}})}]{ArcaSedda2017b}
{Arca-Sedda}, M. \& {Capuzzo-Dolcetta}, R. 2017{\natexlab{b}}, \mnras, 471, 478

\bibitem[{{Arca-Sedda} \&
  {Capuzzo-Dolcetta}(2017{\natexlab{c}})}]{ArcaSedda2017}
{Arca-Sedda}, M. \& {Capuzzo-Dolcetta}, R. 2017{\natexlab{c}}, ArXiv e-prints
  [\eprint[arXiv]{1709.05567}]

\bibitem[{{Arca-Sedda} {et~al.}(2015){Arca-Sedda}, {Capuzzo-Dolcetta},
  {Antonini}, \& {Seth}}]{ArcaSedda2015}
{Arca-Sedda}, M., {Capuzzo-Dolcetta}, R., {Antonini}, F., \& {Seth}, A. 2015,
  \apj, 806, 220

\bibitem[{{Arca-Sedda} {et~al.}(2016){Arca-Sedda}, {Capuzzo-Dolcetta}, \&
  {Spera}}]{ArcaSedda2016}
{Arca-Sedda}, M., {Capuzzo-Dolcetta}, R., \& {Spera}, M. 2016, \mnras, 456,
  2457

\bibitem[{{Arca-Sedda} {et~al.}(2017){Arca-Sedda}, {Kocsis}, \&
  {Brandt}}]{ArcaSedda2017c}
{Arca-Sedda}, M., {Kocsis}, B., \& {Brandt}, T. 2017, ArXiv e-prints
  [\eprint[arXiv]{1709.03119}]

\bibitem[{{Askar} {et~al.}(2018){Askar}, {Arca Sedda}, \&
  {Giersz}}]{ArcaSedda2018b}
{Askar}, A., {Arca Sedda}, M., \& {Giersz}, M. 2018, ArXiv e-prints
  [\eprint[arXiv]{1802.05284}]

\bibitem[{{Askar} {et~al.}(2017){Askar}, {Szkudlarek}, {Gondek-Rosi{\'n}ska},
  {Giersz}, \& {Bulik}}]{Askar2017}
{Askar}, A., {Szkudlarek}, M., {Gondek-Rosi{\'n}ska}, D., {Giersz}, M., \&
  {Bulik}, T. 2017, \mnras, 464, L36

\bibitem[{{Bastian} {et~al.}(2010){Bastian}, {Covey}, \& {Meyer}}]{Bastian2010}
{Bastian}, N., {Covey}, K.~R., \& {Meyer}, M.~R. 2010, \araa, 48, 339

\bibitem[{{Baumgardt} {et~al.}(2018){Baumgardt}, {Amaro-Seoane}, \&
  {Sch{\"o}del}}]{Baumgardt2018}
{Baumgardt}, H., {Amaro-Seoane}, P., \& {Sch{\"o}del}, R. 2018, \aap, 609, A28

\bibitem[{{Bekki}(2007)}]{Bekki2007}
{Bekki}, K. 2007, \pasa, 24, 77

\bibitem[{{Belczynski} {et~al.}(2016){Belczynski}, {Holz}, {Bulik}, \&
  {O'Shaughnessy}}]{Belczynski2016b}
{Belczynski}, K., {Holz}, D.~E., {Bulik}, T., \& {O'Shaughnessy}, R. 2016,
  \nat, 534, 512

\bibitem[{{Belczynski} {et~al.}(2002){Belczynski}, {Kalogera}, \&
  {Bulik}}]{Belczynski2002}
{Belczynski}, K., {Kalogera}, V., \& {Bulik}, T. 2002, \apj, 572, 407

\bibitem[{{Belczynski} {et~al.}(2008){Belczynski}, {Kalogera}, {Rasio}, {Taam},
  {Zezas}, {Bulik}, {Maccarone}, \& {Ivanova}}]{Belczynski2008a}
{Belczynski}, K., {Kalogera}, V., {Rasio}, F.~A., {et~al.} 2008, \apjs, 174,
  223

\bibitem[{{Belczynski} {et~al.}(2017){Belczynski}, {Klencki}, {Meynet},
  {Fryer}, {Brown}, {Chruslinska}, {Gladysz}, {O'Shaughnessy}, {Bulik},
  {Berti}, {Holz}, {Gerosa}, {Giersz}, {Ekstrom}, {Georgy}, {Askar}, {Lasota},
  \& {Wysocki}}]{Belczynski2017b}
{Belczynski}, K., {Klencki}, J., {Meynet}, G., {et~al.} 2017, ArXiv e-prints
  [\eprint[arXiv]{1706.07053}]

\bibitem[{{Belczynski} \& {Taam}(2008)}]{Belczynski2008c}
{Belczynski}, K. \& {Taam}, R.~E. 2008, \apj, 685, 400

\bibitem[{{Belczynski} {et~al.}(2007){Belczynski}, {Taam}, {Kalogera}, {Rasio},
  \& {Bulik}}]{Belczynski2007}
{Belczynski}, K., {Taam}, R.~E., {Kalogera}, V., {Rasio}, F.~A., \& {Bulik}, T.
  2007, \apj, 662, 504

\bibitem[{{Belczynski} {et~al.}(2012){Belczynski}, {Wiktorowicz}, {Fryer},
  {Holz}, \& {Kalogera}}]{Belczynski2012}
{Belczynski}, K., {Wiktorowicz}, G., {Fryer}, C.~L., {Holz}, D.~E., \&
  {Kalogera}, V. 2012, \apj, 757, 91

\bibitem[{{Belloni} {et~al.}(2017{\natexlab{a}}){Belloni}, {Askar}, {Giersz},
  {Kroupa}, \& {Rocha-Pinto}}]{Belloni2017a}
{Belloni}, D., {Askar}, A., {Giersz}, M., {Kroupa}, P., \& {Rocha-Pinto}, H.~J.
  2017{\natexlab{a}}, \mnras, 471, 2812

\bibitem[{{Belloni} {et~al.}(2017{\natexlab{b}}){Belloni}, {Zorotovic},
  {Schreiber}, {Leigh}, {Giersz}, \& {Askar}}]{Belloni2017b}
{Belloni}, D., {Zorotovic}, M., {Schreiber}, M.~R., {et~al.}
  2017{\natexlab{b}}, \mnras, 468, 2429

\bibitem[{{Blanchard} {et~al.}(2017){Blanchard}, {Berger}, {Fong}, {Nicholl},
  {Leja}, {Conroy}, {Alexander}, {Margutti}, {Williams}, {Doctor}, {Chornock},
  {Villar}, {Cowperthwaite}, {Annis}, {Brout}, {Brown}, {Chen}, {Eftekhari},
  {Frieman}, {Holz}, {Metzger}, {Rest}, {Sako}, \&
  {Soares-Santos}}]{Blanchard2017}
{Blanchard}, P.~K., {Berger}, E., {Fong}, W., {et~al.} 2017, \apjl, 848, L22

\bibitem[{{Bottrell} {et~al.}(2017){Bottrell}, {Torrey}, {Simard}, \&
  {Ellison}}]{Bottrell2017}
{Bottrell}, C., {Torrey}, P., {Simard}, L., \& {Ellison}, S.~L. 2017, \mnras,
  467, 2879

\bibitem[{{Breen} \& {Heggie}(2013{\natexlab{a}})}]{Breen2013a}
{Breen}, P.~G. \& {Heggie}, D.~C. 2013{\natexlab{a}}, \mnras, 432, 2779

\bibitem[{{Breen} \& {Heggie}(2013{\natexlab{b}})}]{Breen2013b}
{Breen}, P.~G. \& {Heggie}, D.~C. 2013{\natexlab{b}}, \mnras, 436, 584

\bibitem[{{Capuzzo-Dolcetta}(1993)}]{Capuzzo1993}
{Capuzzo-Dolcetta}, R. 1993, \apj, 415, 616

\bibitem[{{Capuzzo-Dolcetta} \& {Tosta e Melo}(2017)}]{Capuzzo2017}
{Capuzzo-Dolcetta}, R. \& {Tosta e Melo}, I. 2017, \mnras, 472, 4013

\bibitem[{{Chruslinska} {et~al.}(2017){Chruslinska}, {Belczynski}, {Bulik}, \&
  {Gladysz}}]{Chruslinska2017}
{Chruslinska}, M., {Belczynski}, K., {Bulik}, T., \& {Gladysz}, W. 2017,
  \actaa, 67, 37

\bibitem[{{Chruslinska} {et~al.}(2018){Chruslinska}, {Belczynski}, {Klencki},
  \& {Benacquista}}]{Chruslinska2018}
{Chruslinska}, M., {Belczynski}, K., {Klencki}, J., \& {Benacquista}, M. 2018,
  \mnras, 474, 2937

\bibitem[{{Conselice} {et~al.}(2016{\natexlab{a}}){Conselice}, {Wilkinson},
  {Duncan}, \& {Mortlock}}]{Conselice2006}
{Conselice}, C.~J., {Wilkinson}, A., {Duncan}, K., \& {Mortlock}, A.
  2016{\natexlab{a}}, \apj, 830, 83

\bibitem[{{Conselice} {et~al.}(2016{\natexlab{b}}){Conselice}, {Wilkinson},
  {Duncan}, \& {Mortlock}}]{Conselice2016}
{Conselice}, C.~J., {Wilkinson}, A., {Duncan}, K., \& {Mortlock}, A.
  2016{\natexlab{b}}, \apj, 830, 83

\bibitem[{{C{\^o}t{\'e}} {et~al.}(2017){C{\^o}t{\'e}}, {Fryer}, {Belczynski},
  {Korobkin}, {Chru{\'s}li{\'n}ska}, {Vassh}, {Mumpower}, {Lippuner},
  {Sprouse}, {Surman}, \& {Wollaeger}}]{Cote2018}
{C{\^o}t{\'e}}, B., {Fryer}, C.~L., {Belczynski}, K., {et~al.} 2017, ArXiv
  e-prints [\eprint[arXiv]{1710.05875}]

\bibitem[{{C{\^o}t{\'e}} {et~al.}(2006){C{\^o}t{\'e}}, {Piatek}, {Ferrarese},
  {Jord{\'a}n}, {Merritt}, {Peng}, {Ha{\c s}egan}, {Blakeslee}, {Mei}, {West},
  {Milosavljevi{\'c}}, \& {Tonry}}]{Cote2006}
{C{\^o}t{\'e}}, P., {Piatek}, S., {Ferrarese}, L., {et~al.} 2006, \apjs, 165,
  57

\bibitem[{{de Mink} {et~al.}(2009){de Mink}, {Cantiello}, {Langer}, {Pols},
  {Brott}, \& {Yoon}}]{deMink2009}
{de Mink}, S.~E., {Cantiello}, M., {Langer}, N., {et~al.} 2009, \aap, 497, 243

\bibitem[{{de Mink} \& {Mandel}(2016)}]{deMink2016}
{de Mink}, S.~E. \& {Mandel}, I. 2016, \mnras, 460, 3545

\bibitem[{{Dehnen}(1993)}]{Dehnen1993}
{Dehnen}, W. 1993, \mnras, 265, 250

\bibitem[{{Dessart} {et~al.}(2006){Dessart}, {Burrows}, {Ott}, {Livne}, {Yoon},
  \& {Langer}}]{Dessart2006}
{Dessart}, L., {Burrows}, A., {Ott}, C.~D., {et~al.} 2006, \apj, 644, 1063

\bibitem[{{Dominik} {et~al.}(2012){Dominik}, {Belczynski}, {Fryer}, {Holz},
  {Berti}, {Bulik}, {Mandel}, \& {O'Shaughnessy}}]{Dominik2012}
{Dominik}, M., {Belczynski}, K., {Fryer}, C., {et~al.} 2012, \apj, 759, 52

\bibitem[{{Ebrov{\'a}} \& {B{\'{\i}}lek}(2018)}]{Ebrova2018}
{Ebrov{\'a}}, I. \& {B{\'{\i}}lek}, M. 2018, ArXiv e-prints
  [\eprint[arXiv]{1801.01493}]

\bibitem[{{Eldridge} \& {Stanway}(2016)}]{Eldridge2016}
{Eldridge}, J.~J. \& {Stanway}, E.~R. 2016, \mnras, 462, 3302

\bibitem[{{Ferrarese} {et~al.}(2006){Ferrarese}, {C{\^o}t{\'e}}, {Dalla
  Bont{\`a}}, {Peng}, {Merritt}, {Jord{\'a}n}, {Blakeslee}, {Ha{\c s}egan},
  {Mei}, {Piatek}, {Tonry}, \& {West}}]{Ferrarese2006}
{Ferrarese}, L., {C{\^o}t{\'e}}, P., {Dalla Bont{\`a}}, E., {et~al.} 2006,
  \apjl, 644, L21

\bibitem[{{Fregeau} {et~al.}(2004){Fregeau}, {Cheung}, {Portegies Zwart}, \&
  {Rasio}}]{Fregeau2004}
{Fregeau}, J.~M., {Cheung}, P., {Portegies Zwart}, S.~F., \& {Rasio}, F.~A.
  2004, \mnras, 352, 1

\bibitem[{{Fryer} {et~al.}(2012){Fryer}, {Belczynski}, {Wiktorowicz},
  {Dominik}, {Kalogera}, \& {Holz}}]{Fryer2012}
{Fryer}, C.~L., {Belczynski}, K., {Wiktorowicz}, G., {et~al.} 2012, \apj, 749,
  91

\bibitem[{{Fryer} \& {Kusenko}(2006)}]{Fryer2006b}
{Fryer}, C.~L. \& {Kusenko}, A. 2006, \apjs, 163, 335

\bibitem[{{Gallazzi} {et~al.}(2006){Gallazzi}, {Charlot}, {Brinchmann}, \&
  {White}}]{Gallazzi2006}
{Gallazzi}, A., {Charlot}, S., {Brinchmann}, J., \& {White}, S.~D.~M. 2006,
  \mnras, 370, 1106

\bibitem[{{Genel} {et~al.}(2014){Genel}, {Vogelsberger}, {Springel}, {Sijacki},
  {Nelson}, {Snyder}, {Rodriguez-Gomez}, {Torrey}, \& {Hernquist}}]{Genel2014}
{Genel}, S., {Vogelsberger}, M., {Springel}, V., {et~al.} 2014, \mnras, 445,
  175

\bibitem[{{Georgiev} {et~al.}(2016){Georgiev}, {B{\"o}ker}, {Leigh},
  {L{\"u}tzgendorf}, \& {Neumayer}}]{Georgiev2016}
{Georgiev}, I.~Y., {B{\"o}ker}, T., {Leigh}, N., {L{\"u}tzgendorf}, N., \&
  {Neumayer}, N. 2016, \mnras, 457, 2122

\bibitem[{{Giersz}(1998)}]{Giersz1998}
{Giersz}, M. 1998, \mnras, 298, 1239

\bibitem[{{Giersz} {et~al.}(2013){Giersz}, {Heggie}, {Hurley}, \&
  {Hypki}}]{Giersz2013}
{Giersz}, M., {Heggie}, D.~C., {Hurley}, J.~R., \& {Hypki}, A. 2013, \mnras,
  431, 2184

\bibitem[{{Giersz} {et~al.}(2015){Giersz}, {Leigh}, {Hypki}, {L{\"u}tzgendorf},
  \& {Askar}}]{Giersz2015}
{Giersz}, M., {Leigh}, N., {Hypki}, A., {L{\"u}tzgendorf}, N., \& {Askar}, A.
  2015, \mnras, 454, 3150

\bibitem[{{Gnedin} {et~al.}(2014){Gnedin}, {Ostriker}, \&
  {Tremaine}}]{Gnedin2014}
{Gnedin}, O.~Y., {Ostriker}, J.~P., \& {Tremaine}, S. 2014, \apj, 785, 71

\bibitem[{{Graham}(2012)}]{Graham2012}
{Graham}, A.~W. 2012, \mnras, 422, 1586

\bibitem[{{Hamann} \& {Koesterke}(1998)}]{Hamann1998}
{Hamann}, W.-R. \& {Koesterke}, L. 1998, \aap, 335, 1003

\bibitem[{{Harris} {et~al.}(2015){Harris}, {Harris}, \& {Hudson}}]{Harris2015}
{Harris}, W.~E., {Harris}, G.~L., \& {Hudson}, M.~J. 2015, \apj, 806, 36

\bibitem[{{Harris} {et~al.}(2014){Harris}, {Morningstar}, {Gnedin},
  {O'Halloran}, {Blakeslee}, {Whitmore}, {C{\^o}t{\'e}}, {Geisler}, {Peng},
  {Bailin}, {Rothberg}, {Cockcroft}, \& {Barber DeGraaff}}]{Harris2014}
{Harris}, W.~E., {Morningstar}, W., {Gnedin}, O.~Y., {et~al.} 2014, \apj, 797,
  128

\bibitem[{{Heggie} \& {Giersz}(2014)}]{Heggie2014}
{Heggie}, D.~C. \& {Giersz}, M. 2014, \mnras, 439, 2459

\bibitem[{{H{\'e}non}(1971)}]{Henon1971}
{H{\'e}non}, M.~H. 1971, \apss, 14, 151

\bibitem[{{Hoang} {et~al.}(2017){Hoang}, {Naoz}, {Kocsis}, {Rasio}, \&
  {Dosopoulou}}]{Hoang2017}
{Hoang}, B.-M., {Naoz}, S., {Kocsis}, B., {Rasio}, F.~A., \& {Dosopoulou}, F.
  2017, ArXiv e-prints [\eprint[arXiv]{1706.09896}]

\bibitem[{{Hobbs} {et~al.}(2005){Hobbs}, {Lorimer}, {Lyne}, \&
  {Kramer}}]{Hobbs2005}
{Hobbs}, G., {Lorimer}, D.~R., {Lyne}, A.~G., \& {Kramer}, M. 2005, \mnras,
  360, 974

\bibitem[{{Hurley} {et~al.}(2000){Hurley}, {Pols}, \& {Tout}}]{Hurley2000}
{Hurley}, J.~R., {Pols}, O.~R., \& {Tout}, C.~A. 2000, \mnras, 315, 543

\bibitem[{{Hurley} {et~al.}(2002){Hurley}, {Tout}, \& {Pols}}]{Hurley2002}
{Hurley}, J.~R., {Tout}, C.~A., \& {Pols}, O.~R. 2002, \mnras, 329, 897

\bibitem[{{Hypki} \& {Giersz}(2013)}]{Hypki2013}
{Hypki}, A. \& {Giersz}, M. 2013, \mnras, 429, 1221

\bibitem[{{Janka} \& {Mueller}(1994)}]{Janka1994}
{Janka}, H.-T. \& {Mueller}, E. 1994, \aap, 290, 496

\bibitem[{{Jones} {et~al.}(2013){Jones}, {Hirschi}, {Nomoto}, {Fischer},
  {Timmes}, {Herwig}, {Paxton}, {Toki}, {Suzuki}, {Mart{\'{\i}}nez-Pinedo},
  {Lam}, \& {Bertolli}}]{Jones2013}
{Jones}, S., {Hirschi}, R., {Nomoto}, K., {et~al.} 2013, \apj, 772, 150

\bibitem[{{King}(2003)}]{King2003}
{King}, A. 2003, \apjl, 596, L27

\bibitem[{{King}(1966)}]{King1966}
{King}, I.~R. 1966, \aj, 71, 64

\bibitem[{{Kobulnicky} {et~al.}(2014){Kobulnicky}, {Kiminki}, {Lundquist},
  {Burke}, {Chapman}, {Keller}, {Lester}, {Rolen}, {Topel}, {Bhattacharjee},
  {Smullen}, {Vargas {\'A}lvarez}, {Runnoe}, {Dale}, \&
  {Brotherton}}]{Kobulnicky2014}
{Kobulnicky}, H.~A., {Kiminki}, D.~C., {Lundquist}, M.~J., {et~al.} 2014,
  \apjs, 213, 34

\bibitem[{{Kroupa}(1995)}]{Kroupa1995}
{Kroupa}, P. 1995, \mnras, 277 [\eprint{astro-ph/9508084}]

\bibitem[{{Kroupa}(2001)}]{Kroupa2001}
{Kroupa}, P. 2001, \mnras, 322, 231

\bibitem[{{Kruckow} {et~al.}(2018){Kruckow}, {Tauris}, {Langer}, {Kramer}, \&
  {Izzard}}]{Kruckow2018}
{Kruckow}, M.~U., {Tauris}, T.~M., {Langer}, N., {Kramer}, M., \& {Izzard},
  R.~G. 2018, ArXiv e-prints [\eprint[arXiv]{1801.05433}]

\bibitem[{{Lamers} {et~al.}(2010){Lamers}, {Baumgardt}, \&
  {Gieles}}]{Lamers2010}
{Lamers}, H.~J.~G.~L.~M., {Baumgardt}, H., \& {Gieles}, M. 2010, \mnras, 409,
  305

\bibitem[{{Leigh} {et~al.}(2012){Leigh}, {B{\"o}ker}, \& {Knigge}}]{Leigh2012}
{Leigh}, N., {B{\"o}ker}, T., \& {Knigge}, C. 2012, \mnras, 424, 2130

\bibitem[{{Lipunov} {et~al.}(1997){Lipunov}, {Postnov}, \&
  {Prokhorov}}]{Lipunov1997}
{Lipunov}, V.~M., {Postnov}, K.~A., \& {Prokhorov}, M.~E. 1997, Astronomy
  Letters, 23, 492

\bibitem[{{Lyutikov} \& {Toonen}(2017)}]{Lyutikov2017}
{Lyutikov}, M. \& {Toonen}, S. 2017, ArXiv e-prints
  [\eprint[arXiv]{1709.02221}]

\bibitem[{{MacLeod} {et~al.}(2017){MacLeod}, {Antoni}, {Murguia-Berthier},
  {Macias}, \& {Ramirez-Ruiz}}]{MacLeod2017}
{MacLeod}, M., {Antoni}, A., {Murguia-Berthier}, A., {Macias}, P., \&
  {Ramirez-Ruiz}, E. 2017, \apj, 838, 56

\bibitem[{{Maeder}(1987)}]{Maeder1987}
{Maeder}, A. 1987, \aap, 178, 159

\bibitem[{{Mandel} \& {de Mink}(2016)}]{Mandel2016a}
{Mandel}, I. \& {de Mink}, S.~E. 2016, \mnras, 458, 2634

\bibitem[{{Marchant} {et~al.}(2016){Marchant}, {Langer}, {Podsiadlowski},
  {Tauris}, \& {Moriya}}]{Marchant2016}
{Marchant}, P., {Langer}, N., {Podsiadlowski}, P., {Tauris}, T.~M., \&
  {Moriya}, T.~J. 2016, \aap, 588, A50

\bibitem[{{Miyaji} {et~al.}(1980){Miyaji}, {Nomoto}, {Yokoi}, \&
  {Sugimoto}}]{Miyaji1980}
{Miyaji}, S., {Nomoto}, K., {Yokoi}, K., \& {Sugimoto}, D. 1980, \pasj, 32, 303

\bibitem[{{Moriya}(2016)}]{Moriya2016}
{Moriya}, T.~J. 2016, \apjl, 830, L38

\bibitem[{{Morscher} {et~al.}(2015){Morscher}, {Pattabiraman}, {Rodriguez},
  {Rasio}, \& {Umbreit}}]{Morscher2015}
{Morscher}, M., {Pattabiraman}, B., {Rodriguez}, C., {Rasio}, F.~A., \&
  {Umbreit}, S. 2015, \apj, 800, 9

\bibitem[{{Morscher} {et~al.}(2013){Morscher}, {Umbreit}, {Farr}, \&
  {Rasio}}]{Morscher2013}
{Morscher}, M., {Umbreit}, S., {Farr}, W.~M., \& {Rasio}, F.~A. 2013, \apjl,
  763, L15

\bibitem[{{Moustakas} {et~al.}(2013){Moustakas}, {Coil}, {Aird}, {Blanton},
  {Cool}, {Eisenstein}, {Mendez}, {Wong}, {Zhu}, \& {Arnouts}}]{Moustakas2013}
{Moustakas}, J., {Coil}, A.~L., {Aird}, J., {et~al.} 2013, \apj, 767, 50

\bibitem[{{Nayakshin} {et~al.}(2009){Nayakshin}, {Wilkinson}, \&
  {King}}]{Nayakshin2009}
{Nayakshin}, S., {Wilkinson}, M.~I., \& {King}, A. 2009, \mnras, 398, L54

\bibitem[{{Neumayer} \& {Walcher}(2012)}]{Neumayer2012}
{Neumayer}, N. \& {Walcher}, C.~J. 2012, Advances in Astronomy, 2012, 709038

\bibitem[{{Palmese} {et~al.}(2017){Palmese}, {Hartley}, {Tarsitano},
  {Conselice}, {Lahav}, {Allam}, {Annis}, {Lin}, {Soares-Santos}, {Tucker},
  {Brout}, {Banerji}, {Bechtol}, {Diehl}, {Fruchter}, {Garc{\'{\i}}a-Bellido},
  {Herner}, {Levan}, {Li}, {Lidman}, {Misra}, {Sako}, {Scolnic}, {Smith},
  {Abbott}, {Abdalla}, {Benoit-L{\'e}vy}, {Bertin}, {Brooks}, {Buckley-Geer},
  {Carnero Rosell}, {Carrasco Kind}, {Carretero}, {Castander}, {Cunha},
  {D'Andrea}, {da Costa}, {Davis}, {DePoy}, {Desai}, {Dietrich}, {Doel},
  {Drlica-Wagner}, {Eifler}, {Evrard}, {Flaugher}, {Fosalba}, {Frieman},
  {Gaztanaga}, {Gerdes}, {Giannantonio}, {Gruen}, {Gruendl}, {Gschwend},
  {Gutierrez}, {Honscheid}, {Jain}, {James}, {Jeltema}, {Johnson}, {Johnson},
  {Krause}, {Kron}, {Kuehn}, {Kuhlmann}, {Kuropatkin}, {Lima}, {Maia}, {March},
  {Marshall}, {McMahon}, {Menanteau}, {Miller}, {Miquel}, {Neilsen}, {Ogando},
  {Plazas}, {Reil}, {Romer}, {Sanchez}, {Schindler}, {Smith}, {Sobreira},
  {Suchyta}, {Swanson}, {Tarle}, {Thomas}, {Thomas}, {Walker}, {Weller},
  {Zhang}, \& {Zuntz}}]{Palmese2017}
{Palmese}, A., {Hartley}, W., {Tarsitano}, F., {et~al.} 2017, \apjl, 849, L34

\bibitem[{{Park} {et~al.}(2017){Park}, {Kim}, {Lee}, {Bae}, \&
  {Belczynski}}]{Park2017}
{Park}, D., {Kim}, C., {Lee}, H.~M., {Bae}, Y.-B., \& {Belczynski}, K. 2017,
  \mnras, 469, 4665

\bibitem[{{Pavlovskii} \& {Ivanova}(2015)}]{Pavlovskii2015}
{Pavlovskii}, K. \& {Ivanova}, N. 2015, \mnras, 449, 4415

\bibitem[{{Peters}(1964)}]{Peters1964}
{Peters}, P.~C. 1964, Physical Review, 136, 1224

\bibitem[{{Petts} {et~al.}(2015){Petts}, {Gualandris}, \& {Read}}]{Petts2015}
{Petts}, J.~A., {Gualandris}, A., \& {Read}, J.~I. 2015, \mnras, 454, 3778

\bibitem[{{Petts} {et~al.}(2016){Petts}, {Read}, \& {Gualandris}}]{Petts2016}
{Petts}, J.~A., {Read}, J.~I., \& {Gualandris}, A. 2016, \mnras, 463, 858

\bibitem[{{Peuten} {et~al.}(2016){Peuten}, {Zocchi}, {Gieles}, {Gualandris}, \&
  {H{\'e}nault-Brunet}}]{Peuten2016}
{Peuten}, M., {Zocchi}, A., {Gieles}, M., {Gualandris}, A., \&
  {H{\'e}nault-Brunet}, V. 2016, \mnras, 462, 2333

\bibitem[{{Podsiadlowski} {et~al.}(2004){Podsiadlowski}, {Langer},
  {Poelarends}, {Rappaport}, {Heger}, \& {Pfahl}}]{Podsiadlowski2004}
{Podsiadlowski}, P., {Langer}, N., {Poelarends}, A.~J.~T., {et~al.} 2004, \apj,
  612, 1044

\bibitem[{{Portegies Zwart} {et~al.}(2004){Portegies Zwart}, {Baumgardt},
  {Hut}, {Makino}, \& {McMillan}}]{PortegiesZwart2004}
{Portegies Zwart}, S.~F., {Baumgardt}, H., {Hut}, P., {Makino}, J., \&
  {McMillan}, S.~L.~W. 2004, \nat, 428, 724

\bibitem[{{Rodriguez} {et~al.}(2016{\natexlab{a}}){Rodriguez}, {Chatterjee}, \&
  {Rasio}}]{Rodriguez2016b}
{Rodriguez}, C.~L., {Chatterjee}, S., \& {Rasio}, F.~A. 2016{\natexlab{a}},
  \prd, 93, 084029

\bibitem[{{Rodriguez} {et~al.}(2016{\natexlab{b}}){Rodriguez}, {Haster},
  {Chatterjee}, {Kalogera}, \& {Rasio}}]{Rodriguez2016a}
{Rodriguez}, C.~L., {Haster}, C.-J., {Chatterjee}, S., {Kalogera}, V., \&
  {Rasio}, F.~A. 2016{\natexlab{b}}, \apjl, 824, L8

\bibitem[{{Rodriguez} {et~al.}(2015){Rodriguez}, {Morscher}, {Pattabiraman},
  {Chatterjee}, {Haster}, \& {Rasio}}]{Rodriguez2015}
{Rodriguez}, C.~L., {Morscher}, M., {Pattabiraman}, B., {et~al.} 2015, Physical
  Review Letters, 115, 051101

\bibitem[{{Rodriguez} {et~al.}(2016{\natexlab{c}}){Rodriguez}, {Morscher},
  {Wang}, {Chatterjee}, {Rasio}, \& {Spurzem}}]{Rodriguez2016-dr}
{Rodriguez}, C.~L., {Morscher}, M., {Wang}, L., {et~al.} 2016{\natexlab{c}},
  \mnras, 463, 2109

\bibitem[{{Rodriguez} {et~al.}(2016{\natexlab{d}}){Rodriguez}, {Zevin},
  {Pankow}, {Kalogera}, \& {Rasio}}]{Rodriguez2016c}
{Rodriguez}, C.~L., {Zevin}, M., {Pankow}, C., {Kalogera}, V., \& {Rasio},
  F.~A. 2016{\natexlab{d}}, \apjl, 832, L2

\bibitem[{{Rossa} {et~al.}(2006){Rossa}, {van der Marel}, {B{\"o}ker},
  {Gerssen}, {Ho}, {Rix}, {Shields}, \& {Walcher}}]{Rossa2006}
{Rossa}, J., {van der Marel}, R.~P., {B{\"o}ker}, T., {et~al.} 2006, \aj, 132,
  1074

\bibitem[{{Sana} {et~al.}(2012){Sana}, {de Mink}, {de Koter}, {Langer},
  {Evans}, {Gieles}, {Gosset}, {Izzard}, {Le Bouquin}, \&
  {Schneider}}]{Sana2012}
{Sana}, H., {de Mink}, S.~E., {de Koter}, A., {et~al.} 2012, Science, 337, 444

\bibitem[{{Schechter}(1976)}]{Schechter1976}
{Schechter}, P. 1976, \apj, 203, 297

\bibitem[{{Schwab} {et~al.}(2015){Schwab}, {Quataert}, \&
  {Bildsten}}]{Schwab2015}
{Schwab}, J., {Quataert}, E., \& {Bildsten}, L. 2015, \mnras, 453, 1910

\bibitem[{{Scott} \& {Graham}(2013)}]{Scott2013}
{Scott}, N. \& {Graham}, A.~W. 2013, \apj, 763, 76

\bibitem[{{Sippel} \& {Hurley}(2013)}]{Sippel2013}
{Sippel}, A.~C. \& {Hurley}, J.~R. 2013, \mnras, 430, L30

\bibitem[{{Snyder} {et~al.}(2015){Snyder}, {Torrey}, {Lotz}, {Genel},
  {McBride}, {Vogelsberger}, {Pillepich}, {Nelson}, {Sales}, {Sijacki},
  {Hernquist}, \& {Springel}}]{Snyder2015}
{Snyder}, G.~F., {Torrey}, P., {Lotz}, J.~M., {et~al.} 2015, \mnras, 454, 1886

\bibitem[{{Stevenson} {et~al.}(2017){Stevenson}, {Vigna-G{\'o}mez}, {Mandel},
  {Barrett}, {Neijssel}, {Perkins}, \& {de Mink}}]{Stevenson2017}
{Stevenson}, S., {Vigna-G{\'o}mez}, A., {Mandel}, I., {et~al.} 2017, Nature
  Communications, 8, 14906

\bibitem[{{Stodolkiewicz}(1986)}]{Stodolkiewicz1986}
{Stodolkiewicz}, J.~S. 1986, \actaa, 36, 19

\bibitem[{{Torrey} {et~al.}(2015){Torrey}, {Wellons}, {Machado}, {Griffen},
  {Nelson}, {Rodriguez-Gomez}, {McKinnon}, {Pillepich}, {Ma}, {Vogelsberger},
  {Springel}, \& {Hernquist}}]{Torrey2015}
{Torrey}, P., {Wellons}, S., {Machado}, F., {et~al.} 2015, \mnras, 454, 2770

\bibitem[{{Tremaine} {et~al.}(1975){Tremaine}, {Ostriker}, \&
  {Spitzer}}]{Tremaine1975}
{Tremaine}, S.~D., {Ostriker}, J.~P., \& {Spitzer}, Jr., L. 1975, \apj, 196,
  407

\bibitem[{{Troja} {et~al.}(2017){Troja}, {Piro}, {van Eerten}, {Wollaeger},
  {Im}, {Fox}, {Butler}, {Cenko}, {Sakamoto}, {Fryer}, {Ricci}, {Lien}, {Ryan},
  {Korobkin}, {Lee}, {Burgess}, {Lee}, {Watson}, {Choi}, {Covino}, {DAvanzo},
  {Fontes}, {Gonzalez}, {Khandrika}, {Kim}, {Kim}, {Lee}, {Lee}, {Kutyrev},
  {Lim}, {Sanchez-Ramirez}, {Veilleux}, {Wieringa}, \& {Yoon}}]{Troja2017}
{Troja}, E., {Piro}, L., {van Eerten}, H., {et~al.} 2017, \nat, 551, 71

\bibitem[{{Tsatsi} {et~al.}(2017){Tsatsi}, {Mastrobuono-Battisti}, {van de
  Ven}, {Perets}, {Bianchini}, \& {Neumayer}}]{Tsatsi2017}
{Tsatsi}, A., {Mastrobuono-Battisti}, A., {van de Ven}, G., {et~al.} 2017,
  \mnras, 464, 3720

\bibitem[{{Turner} {et~al.}(2012){Turner}, {C{\^o}t{\'e}}, {Ferrarese},
  {Jord{\'a}n}, {Blakeslee}, {Mei}, {Peng}, \& {West}}]{Turner2012}
{Turner}, M.~L., {C{\^o}t{\'e}}, P., {Ferrarese}, L., {et~al.} 2012, \apjs,
  203, 5

\bibitem[{{Tutukov} \& {Yungelson}(1993)}]{Tutukov1993}
{Tutukov}, A.~V. \& {Yungelson}, L.~R. 1993, \mnras, 260, 675

\bibitem[{{Vink} \& {de Koter}(2005)}]{Vink2005}
{Vink}, J.~S. \& {de Koter}, A. 2005, \aap, 442, 587

\bibitem[{{Vink} {et~al.}(2001){Vink}, {de Koter}, \& {Lamers}}]{Vink2001}
{Vink}, J.~S., {de Koter}, A., \& {Lamers}, H.~J.~G.~L.~M. 2001, \aap, 369, 574

\bibitem[{{Vogelsberger} {et~al.}(2014{\natexlab{a}}){Vogelsberger}, {Genel},
  {Springel}, {Torrey}, {Sijacki}, {Xu}, {Snyder}, {Bird}, {Nelson}, \&
  {Hernquist}}]{Vogelsberger2014a}
{Vogelsberger}, M., {Genel}, S., {Springel}, V., {et~al.} 2014{\natexlab{a}},
  \nat, 509, 177

\bibitem[{{Vogelsberger} {et~al.}(2014{\natexlab{b}}){Vogelsberger}, {Genel},
  {Springel}, {Torrey}, {Sijacki}, {Xu}, {Snyder}, {Nelson}, \&
  {Hernquist}}]{Vogelsberger2014}
{Vogelsberger}, M., {Genel}, S., {Springel}, V., {et~al.} 2014{\natexlab{b}},
  \mnras, 444, 1518

\bibitem[{{Voss} \& {Tauris}(2003)}]{Voss2003}
{Voss}, R. \& {Tauris}, T.~M. 2003, \mnras, 342, 1169

\bibitem[{{Wang} {et~al.}(2016){Wang}, {Spurzem}, {Aarseth}, {Giersz}, {Askar},
  {Berczik}, {Naab}, {Schadow}, \& {Kouwenhoven}}]{Wang2016}
{Wang}, L., {Spurzem}, R., {Aarseth}, S., {et~al.} 2016, \mnras, 458, 1450

\bibitem[{{Webb} \& {Leigh}(2015)}]{Webb2015}
{Webb}, J.~J. \& {Leigh}, N.~W.~C. 2015, \mnras, 453, 3278

\bibitem[{{Webbink}(1984)}]{Webbink1984}
{Webbink}, R.~F. 1984, \apj, 277, 355

\bibitem[{{Woosley}(2016)}]{Woosley2016}
{Woosley}, S.~E. 2016, \apjl, 824, L10

\bibitem[{{Xu} \& {Li}(2010)}]{Xu2010}
{Xu}, X.-J. \& {Li}, X.-D. 2010, \apj, 722, 1985

\bibitem[{{Yoon} \& {Langer}(2005)}]{Yoon2005}
{Yoon}, S.-C. \& {Langer}, N. 2005, \aap, 443, 643

\bibitem[{{Yoon} {et~al.}(2006){Yoon}, {Langer}, \& {Norman}}]{Yoon2006}
{Yoon}, S.-C., {Langer}, N., \& {Norman}, C. 2006, \aap, 460, 199

\end{thebibliography}

\begin{appendix}

\section{The Illustris Simulation}
\label{sec.app0}

Among all the galaxies in the local Universe, only $1/3$ is found in 
ellipticals \citep{Conselice2006} with ages spanning the range $1-10$ Gyr 
\citep{Gallazzi2006}, a feature that is represented reasonably well in the 
latest generation of cosmological simulations, such as the Illustris 
simulation \citep{Vogelsberger2014,Snyder2015}, although recently it has 
been argued that it overproduces disc galaxies with stellar mass in the
range $10^{10}-10^{11}\msun$ \citep{Bottrell2017}.

There is some evidence that the stellar mass in low star formation rate 
galaxies is $\sim 52\%$ the total stellar mass in galaxies
~\citep{Moustakas2013,Vogelsberger2014a}. If we assume that ellipticals 
dominate the population of galaxies with little or no star formation, then 
our adopted estimate of stellar mass contained in ellipticals ($1/3$) may 
change to $1/2$. This would slightly (by a factor of $1.5$) increase all 
the NS-NS merger rates calculated in our study. However, such a change has 
no influence on our conclusions. 

We take advantage of the Illustris-1 simulation which modeled the evolution 
of a cosmological cube with side $L = 106.5$ Mpc using $1.8\times 10^{10}$ 
particles for representing baryonic and dark matter and full physics 
prescriptions as described in \cite{Vogelsberger2014}.

We calculated the total number of objects having a stellar mass above 
$10^6\msun$  at redshift $z=0$, in order to take into account all the 
bounded stellar systems available in the simulation. This led to a total 
number of objects $N_{\rm Ill} = 238,525$ with total stellar mass 
$M_{\rm Ill} = 3.9 \times 10^{14} \msun$.

Then, we rescaled the total number and stellar mass of all the 
elliptical galaxies ($1/3$ of the total) contained within 
$100$ Mpc$^3$, scaling the quantities above as:
\begin{eqnarray}
N_{\rm ell,tot} =&
\displaystyle\frac{1}{3}\left(\frac{100}{106.5}\right)^3 
N_{\rm Ill}&= 65,821\\
M_{\rm ell,tot} =& 
\displaystyle\frac{1}{3}\left(\frac{100}{106.5}\right)^3
M_{\rm Ill}&= 1.1\times 10^{14}\msun.
\end{eqnarray}

\end{appendix}

\end{document}